# Using Optimization to Solve Positive LPs Faster in Parallel

Zeyuan Allen-Zhu
zeyuan@csail.mit.edu
MIT CSAIL

Lorenzo Orecchia
orecchia@mit.edu
MIT Math

July 7, 2014[*]

**Abstract**

Positive linear programs (LP), also known as packing and covering linear programs, are an important class of problems that bridges computer science, operations research, and optimization. Despite the consistent efforts on this problem, all known nearly-linear-time algorithms require $\widetilde{O}(\varepsilon^{-4})$ iterations to converge to $1 \pm \varepsilon$ approximate solutions. This $\varepsilon^{-4}$ dependence has not been improved since 1993, and limits the performance of parallel implementations for such algorithms. Moreover, previous algorithms and their analyses rely on update steps and convergence arguments that are combinatorial in nature and do not seem to arise naturally from an optimization viewpoint.

In this paper, we leverage new insights from optimization theory to construct a novel algorithm that breaks the longstanding $\varepsilon^{-4}$ barrier. Our algorithm has a simple analysis and a clear motivation. Our work introduces a number of novel techniques, such as the combined application of gradient descent and mirror descent, and a truncated, smoothed version of the standard multiplicative weight update, which may be of independent interest.

---

[*]First version appeared on this date. This newer version contains polished writing.

# 1 Introduction

Fractional packing and covering linear programs (LP) are described with non-negative matrices, non-negative constraints, and non-negative variables. They are also known as positive linear programs as originally studied by Luby and Nisan [LN93].

A generic packing LP takes the form $\max\{c^T x \,:\, Ax \leq b\}$ where $c \in \mathbb{R}^n_{\geq 0}$, $b \in \mathbb{R}^m_{\geq 0}$, and $A \in \mathbb{R}^{m \times n}_{\geq 0}$; similarly, a covering LP can be written as $\min\{b^T y \,:\, A^T y \geq c\}$, with the same requirements on $A, b$, and $c$. As in other works, we assume without loss of generality that the LP is in its *standard form*: $b = \mathbb{1}$ and $c = \mathbb{1}$:[1]

$$\text{Packing LP:} \quad \max_{x \geq 0}\{\mathbb{1}^T x \,:\, Ax \leq \mathbb{1}\} \,, \tag{1.1}$$

$$\text{Covering LP:} \quad \min_{y \geq 0}\{\mathbb{1}^T y \,:\, A^T y \geq \mathbb{1}\} \,. \tag{1.2}$$

Since the two programs are dual to each other, we denote by OPT their shared optimal value. We say that $x$ is a $(1-\varepsilon)$-approximation for the packing LP if $Ax \leq \mathbb{1}$ and $\mathbb{1}^T x \geq (1-\varepsilon)\mathsf{OPT}$, and $y$ a $(1+\varepsilon)$-approximation for the covering LP if $A^T y \geq \mathbb{1}$ and $\mathbb{1}^T y \leq (1+\varepsilon)\mathsf{OPT}$.

Of course, it is possible to adopt the general Interior Point or Ellipsoid Methods to obtain approximate solvers with a $\log(1/\varepsilon)$ dependence on the number of iterations. However, the computational cost of such algorithms is typically very high, as each iteration requires the solution of a system of linear equations in $A^T A$. As a consequence, this approach is simply not suitable to the solution of large-scale problems.

To address this issue, researchers have developed iterative *approximate* solvers that achieve a better dependence on the problem size at the cost of having a $\mathsf{poly}(1/\varepsilon)$ dependence on the approximation parameter $\varepsilon$. These algorithms rely crucially on the power of *multiplicative weight update methods* (see the survey by Arora, Hazan and Kale [AHK12]). Multiplicative weight update methods can be viewed as special cases of the *mirror descent* method, a widely-used first-order method in optimization (see for instance [AO17] for this relationship). Such methods achieve fast running times by eschewing any structure in the problem and only accessing the instance in a restricted, quick fashion through the computation of gradients of the objective.

As a result, iterative approximate solvers often require a larger number of iterations, i.e., one that depends on $\mathsf{poly}(1/\varepsilon)$, but each iteration consists only of a small number of simple steps (such as matrix-vector multiplications or sorting operations) and requires only nearly-linear work in $N$ and $O(\log N)$ depth, even in the weak EREW model of the Parallel Random Access Machine (PRAM).

Such fast approximate positive-LP solvers have been widely used in approximation algorithms (e.g., MINSETCOVER [LN93], MAXSET, MAXDICUT, MAX-$k$-CSP [Tre98], bipartite matching), probabilistic checkable proofs [Tre98], zero-sum matrix games [Nes05], scheduling [PST95], graph embedding [PST95], flow controls [BBR97, BBR04], auction mechanisms [ZN01], wireless sensor networks [BN00], and many other areas. In addition, techniques developed in this line of research have also inspired many other important results, most notably regarding fast algorithms for multi-commodity flow problems [PST95, Fle00, GK07, Mad10, AKR12].

Previous approximate solvers can be further divided into two classes.

---

[1] This can be achieved simply by scaling.

| Problem | Paper | Total Work | Number of Iterations[a] | Notes |
|---|---|---|---|---|
| p/c LP | [LN93] | $\frac{\log^2 N}{\varepsilon^4} \times (N \log n)$ | $\frac{\log^2 N}{\varepsilon^4}$ | |
| p/c LP | [BBR97, BBR04] | $\frac{\log^3 N}{\varepsilon^4} \times N$ | $\frac{\log^3 N}{\varepsilon^4}$ | |
| p/c LP | [You01] | $\frac{\log^3 N}{\varepsilon^4} \times N$ | $\frac{\log^3 N}{\varepsilon^4}$ | mixed p/c |
| p/c LP | [AK08a] | $\frac{\log^4 N}{\varepsilon^5} \times N$ | $\frac{\log^4 N}{\varepsilon^5}$ | stateless |
| p/c LP | **[this paper]** | $\frac{\log^2 N}{\varepsilon^3} \times N$ | $\frac{\log^2 N}{\varepsilon^3}$ | semi-stateless |
| p/c LP | [You01] | $\frac{\log N}{\varepsilon^2} \times (md + N)$ [b] | $\frac{\log N}{\varepsilon^2} \times (n+m)$ | not parallelizable |
| p/c LP | [You14] | $\frac{\log N}{\varepsilon^2} \times N$ | $\frac{\log N}{\varepsilon^2} \times (n+m)$ | not parallelizable |
| p/c LP | [KY13] | $\frac{\log N}{\varepsilon^2} \times (n+m) + N$ | $\frac{\log N}{\varepsilon^2} \times (n+m)$ | not parallelizable |

Table 1: Comparisons among width-independent approximate solvers for positive LPs.

---

[a] For most parallelizable solvers, an iteration is dominated by a matrix-vector multiplicative that can be implemented in $O(N)$ total work. However, an iteration of Luby-Nisan is more complicated, and to the best of our knowledge, we only know how to implement it in $O(nm)$ or $O(N \log n)$ total work, rather than $O(N)$.

[b] $d$ is the maximum number of constraints each variable is in; $md$ may be larger than $N$.

**Width-Dependent Solvers.** These algorithms[2] require a number of iterations that is at least linearly dependent on $\rho \cdot \mathsf{OPT}$, where $\rho$ is the largest entry, i.e. the *width*, of matrix $A$. Since $\mathsf{OPT} \geq 1/\rho$, this value $\rho \cdot \mathsf{OPT}$ is at least 1. However, since $\mathsf{OPT}$ can easily be as large as 1 or even more than $n$, the resulting running time is not polynomial, but only pseudo-polynomial. In particular, positive LPs can be solved in $O(\frac{\rho^2 \mathsf{OPT}^2 \log m}{\varepsilon^2})$ iterations [PST95], or $O(\frac{\rho \mathsf{OPT} \log m}{\varepsilon^2})$ iterations using negative-width techniques [AHK12]. These algorithms strongly rely on multiplicative weight updates and only require "oracle-access" to the matrix $A$.

When $A$ is given explicitly like in this paper, the number of iterations can be reduced to $O(\frac{\rho \mathsf{OPT} \log m}{\varepsilon})$ by deploying more advanced optimization tools such as Nesterov's accelerated gradient method [Nes05], or Nemirovski's mirror prox method [Nem04]. It is also worth noting that Bienstock and Iyengar [BI04] have converted this dependence on $\rho \mathsf{OPT}$ into a more benign, yet linear dependence on $n$. More specifically, their iteration count is $O(\varepsilon^{-1}\sqrt{Kn \log m})$ where $K$ is the maximum number of non-zeros per row of $A$. This is $O(\varepsilon^{-1} n \sqrt{\log m})$ in the worst case.

**Width-Independent Solvers.** In this paper, we are interested in a second, more efficient class of methods, i.e. *width-independent*,[3] truly polynomial-time approximate solvers (see Table 1).

This line of research was initiated by a seminal paper of Luby and Nisan [LN93], who were able to remove the dependence from the width and give an algorithm running in $O\left(\frac{\log^2 N}{\varepsilon^4}\right)$ iterations. Theirs is the first *nearly-linear-time* approximate solver for positive LPs and also the first to run in parallel in nearly-linear-work and polylogarithmic depth. This algorithm was later simplified

---

[2] Note that most width-dependent solvers are studied under the minmax form of positive LPs:

$$\min_{\substack{x \geq 0 \\ \mathbb{1}^T x = 1}} \max_{\substack{y \geq 0 \\ \mathbb{1}^T y = 1}} y^T A x \ ,$$

whose optimal value equals $1/\mathsf{OPT}$. Their approximation guarantees are often written in terms of the *additive* error. We have translated their performances to the multiplicative error for a fair comparison.

[3] Some of these solvers may still have a $\mathsf{polylog}(\rho)$ dependence. Since each occurrence of $\log(\rho)$ can typically be replaced with $\log(nm)$ after slightly modifying the instance matrix $A$, we have done so in Table 1 for a fair comparisons.

and made explicit for parallelization by Bartal, Byers and Raz [BBR97], improved to allow mixed packing and covering by Young [You01], and generalized by Awerbuch and Khandekar [AK08a] to the computational model where processors are restricted to be 'stateless'. These solvers are *parallelizable* because they only require $O(\text{polylog}(N)/\varepsilon^{O(1)})$ iterations to converge to $1 \pm \varepsilon$ approximate solutions. They are nearly-linear time because each iteration runs in nearly-linear time.

A separate line of work starting from Bartal, Byers and Raz [BBR97, BBR04] eschews the parallelization constraint to design *sequential* width-independent solvers with a better $\varepsilon$ dependence. At high level, these algorithms modify the candidate LP solutions coordinate by coordinate and therefore require at least a linear number of iterations to converge. For instance, the algorithm of Koufogiannakis and Young [KY13] runs in nearly-linear total time $O(N + \frac{\log N}{\varepsilon^2} \times (n+m))$, but requires $O(\frac{\log N}{\varepsilon^2}(n+m))$ iterations to converge to $1 \pm \varepsilon$ approximate solutions. In contrast, as we shall discuss later in Section 1.1, parallelizable solvers modify all coordinates of the candidate LP solution *at once* per iteration, thus converging in a much smaller polylogarithmic number of iterations. For this reason, the design of parallelizable solvers faces different technical challenges from that of sequential ones, because the update rules are much more restrictive. We have summarized prior results on sequential solvers in Table 1.

To sum up, despite the amount of work in this area, the $O\bigl(\frac{\log^2 N}{\varepsilon^4}\bigr)$-iteration-count has not been improved since the original paper of Luby and Nisan. This lack of progress constitutes a significant limitation, as the $\varepsilon^{-4}$-dependence on the approximation parameter $\varepsilon$ is particularly pour. The question of how to go beyond $\varepsilon^{-4}$ has been raised by Young [You01] and remained open until now. In this paper, we give an answer to this question and provide a brief empirical evaluation supporting the idea that the performance gains achieved by our algorithm in the worst-case actually translate into practice.

## 1.1 Our Results

In this paper, we present an algorithm $\texttt{PosLPSolver}(A, \varepsilon)$ that runs only in $O(\frac{\log n \cdot \log(nm/\varepsilon)}{\varepsilon^3})$ iterations, and each iteration consists mostly of a matrix-vector multiplication so can be implemented in $O(\log N)$ parallel depth. This is a total work of $O\bigl(\frac{\log n \cdot \log(nm/\varepsilon)}{\varepsilon^3} \cdot N\bigr)$. (See a full comparison between our and previous results in Table 1.) Besides being the fastest parallel algorithm for solving positive LPs to date, our method also is surprisingly simple and enjoys a 'semi-stateless' property, i.e. is stateless except for requiring a global clock (see Appendix B).

Our algorithms works by optimizing a relaxation of the original packing LP (see Definition 2.1), where the hard constraint $Ax \leq 1$ is replaced by an exponential penalty function for violating the constraint.[4] This initial step ensures that our candidate iterative solutions remain approximately feasible throughout the evolution of the algorithm. It also leads us to optimize our modified objective by updating our current iterate $x^{(k)}$ using gradient information. This is done by computing a feedback vector $v$ so that $v_i \stackrel{\text{def}}{=} \sum_{j=1}^{m} A_{i,j} \cdot \exp^{\frac{1}{\mu}((Ax)_j - 1)} - 1 \in [-1, \infty)$ for each variable $i \in [n]$, and performing a multiplicative update $x_i \leftarrow x_i \cdot \exp^{-\alpha \cdot \mathbb{T}(v_i)}$. Here, our thresholding function $\mathbb{T}(v) = v$ for $v \in [-1, 1] \setminus [-\varepsilon, \varepsilon]$, $\mathbb{T}(v) = 0$ for $v \in [-\varepsilon, \varepsilon]$, and $\mathbb{T}(v) = 1$ for $v > 1$; and $\alpha = \frac{\varepsilon \mu}{4}$ is some fixed constant.

**Our Techniques.** Our result fundamentally differs from all previous width-independent solvers both in the algorithm specification and in its analysis. Like previous works, we also update the coordinates of $x$ simultaneously and multiplicatively. However, previous methods treat all relevant coordinates alike, multiplying each of them either by $1 + \alpha$ or $1 - \alpha$, for some fixed constant $\alpha$. Instead, our use of the feedback vector $v$ (along with the thresholding function) allows us to update the coordinates by a factor between $e^{\pm\alpha} \approx 1 \pm \alpha$ and $e^{\pm\varepsilon\alpha} \approx 1 \pm \varepsilon\alpha$. This *discriminative multiplicative update* rule is a key step in overcoming the $1/\varepsilon^4$ barrier.

---

[4]This standard technique in optimization is used explicitly in [AK08a] and implicitly in [LN93] and [You01].

More importantly, our work introduces a completely novel way of analyzing the performance of our algorithm. More specifically, previous methods [LN93, BBR97, You01, GK07] fall into the following framework: the method is divided into $\widetilde{\Omega}(\frac{1}{\varepsilon^2})$ *phases*, with each phase having a different parameter setting. Each phase itself consists of $\widetilde{\Omega}(\frac{1}{\varepsilon^2})$ iterations. This immediately prevents their analyses from breaking the $\frac{1}{\varepsilon^4}$ barrier[5].

In contrast, we interpret the packing LP problem as a purely optimization question, i.e., to minimize $f(x)$ for some convex function $f$. Next, in each iteration of the algorithm, we interpret the feedback vector $v$ as the gradient $\nabla f(x) \in [-1, \infty)^n$, and divide it into two components, the large component $\eta \in [0, \infty)^n$ and the small (and truncated) component $\xi \in [-1, 1]^n$, satisfying $\nabla f(x) \approx \eta + \xi$. The key observation now is to interpret our update $x_i \leftarrow x_i \cdot \exp^{-\alpha \cdot \mathbb{T}(v_i)}$ as performing two different kind of steps at the same time:

- a *gradient descent*[6] step (on $\eta$), to ensure that $f(x)$ decreases by a large amount at each step; and
- a *mirror descent* step (on $\xi$), to ensures that the average *regret* of the history of the steps is small.

Both gradient and mirror descent are well-known tools from optimization (see for instance [Nes04, BN13] and, for starters, mirror descent is a generalization of multiplicative weight updates). This 'duality' view allows us to combine the analysis of both gradient and mirror descent for a faster algorithm, and is the key to bypass the combinatorial/phaseful analysis used by all previous results. More generally, the same authors of this paper observed that gradient and mirror descent have complementary performances, and *coupling* these two methods often leads to better running times [AO17].

We develop two more techniques that may be of independent interests, one for the gradient descent analysis and one for the mirror descent analysis. In our gradient descent view, since $f(x)$ does not satisfy any Lipschitz gradient property, the classical convergence analysis of gradient descent (see [Nes04]) no longer applies.[7] Instead, we adopt a *multiplicative Lipschitz gradient property*: if each coordinate of $x$ changes multiplicatively by a little, the gradient does not change too much multiplicatively as well. This property enables us to produce a promise on the decrease of the objective $f(x)$ in each step.

In our mirror descent analysis, we have developed a *gradient truncation* technique that removes large components from the gradient, delegating their contribution to the gradient descent analysis. This effectively reduces the width experienced by our mirror descent algorithm.

Finally, we emphasize that our optimization view for solving positive LPs should be seen as yet another example on *designing combinatorial algorithms based on insights from optimization*. Before our work, the updates on $x$ are maximally aggressive, since they arise naturally from a combinatorial approach to the solution of the original LP program. In our algorithm, we have smoothed out the updates on $x$ so that, for coordinates whose absolute feedbacks $|v_i|$ are small, we perform less aggressive steps. While one may find such intuition very legitimate, without the optimization interpretation behind it, it is very hard to analyze the resulting algorithm or even to find the *right* step length. For instance, the algorithm of [AK08a] is similar to ours in terms of the updates on $x$. However, the simple difference between the choices of step length makes our algorithm faster than theirs, $\log^2 N/\varepsilon^3$ vs. $\log^4 N/\varepsilon^5$. Moreover, our step lengths are in fact *less aggressive* than theirs in terms of decreasing the objective $f(x)$. We also provide an empirical evaluation in Appendix A to support this comparison.

---

[5] Although the algorithm in [AK08a] does not explicitly require phases, its convergence analysis divides the iterations into $\Omega(\frac{\log^2 N}{\varepsilon^3})$ phases each with $\Omega(\frac{\log^2 N}{\varepsilon^2})$ iterations.

[6] It is important to note here that we have generalized the notion of "gradient descent" to indicate any descent step that is guaranteed to decrease the objective. This is in contrast to mirror descent, that does not necessarily decrease the objective at each iteration.

[7] The Lipschitz gradient property (also known as Lipschitz smooth property in the literature) says that $\|\nabla f(x_1) - \nabla f(x_2)\| \leq L \cdot \|x_1 - x_2\|$ for some constant $L$ and some special choice of norm. If one forces $f(x)$ to satisfy this property, the algorithm falls into the category of [Nes05] and becomes width-dependent.

**The Stateless Feature.** Some parallelizable algorithms enjoy a desirable *stateless* feature. Informally, this feature requires that the updates of each processor only depend on the current feedback, and not on the history or on any global variable. The only known stateless solver for positive LPs is due to Awerbuch and Khandekar [AK08a], but their method is much slower than that of Luby and Nisan (see Table 1). Stateless algorithms enjoy a number of features (P1) *self-stabilization*, (P2) *robustness against incremental adjustments*, and (P3) *no global clock*. We point out that our algorithm is 'semi-stateless' (introduced in Appendix B): that is, it exhibits properties (P1) and (P2). Unfortunately, our current proof technique requires the use of a global clock for the parallelized algorithm. Instead, [AK08a] only requires that the desired number of iterations are performed synchronously with the global clock, while between consecutive iterations each processor can run on its own arbitrarily without synchronization.

## 1.2 Roadmap

We transfer the positive LP problem into an optimization question in Section 2, provide our packing LP solver in Section 3, and turn the same algorithm into a covering LP solver in Section 4. We also provide a brief empirical evaluation comparing the performance of our algorithm against previous ones in Appendix A. We defer the argument of the semi-statelessness of our LP solver to Appendix B. Some missing proofs are included in the appendix.

## 2 Smoothing the Positive LP Objective

In this section we introduce the smoothed objective $f_\mu(x)$ that we are going to minimize in order to approximately solve the packing LP, by turning each row of the LP constraint $Ax \leq \mathbb{1}$ into an exponential penalty function so that we only need to require $x \geq 0$ throughout the algorithm.

Let $x^*$ be any optimal solution of the packing LP (1.1). Throughout this paper, we use indices $i \in [n]$ for the columns of $A$, and $j \in [m]$ for the rows of $A$. We denote by $A_{:i}$ the $i$-th column vector of $A$, and $A_{j:}$ the $j$-th row vector of $A$. We assume without loss of generality that

$$\min_{i \in [n]}\{\|A_{:i}\|_\infty\} = 1 \;, \tag{2.1}$$

since otherwise one can scale $A$ by a constant factor, and the solution OPT as well as $x^*$ are only affected by this same constant factor.

We now introduce our smoothed objective $f_\mu(x)$.

**Definition 2.1.** *Letting parameter $\mu \stackrel{\text{def}}{=} \frac{\varepsilon}{4\log(nm/\varepsilon)}$, we define the smoothed objective $f_\mu(x)$ as*

$$f_\mu(x) \stackrel{\text{def}}{=} \mu\sum_{j=1}^m \exp^{\frac{1}{\mu}((Ax)_j - 1)} - \mathbb{1}^T x \;.$$

We wish to study the *minimization* problem on $f_\mu(x)$, subject to the constraint that each coordinate $x_i \geq 0$ is non-negative. We denote by $x \geq 0$ this positive orthant.

Intuitively this objective $f_\mu(x)$ should capture the original packing LP (1.1) approximately as follows. On one hand, we want to maximize $\mathbb{1}^T x$ so the negative term $-\mathbb{1}^T x$ shows up in $f_\mu(x)$. On the other, if $(Ax)_j \geq 1 + \varepsilon$ for some $j$, the exponential penalty in $f_\mu(x)$ introduces a value that is at least $\exp^{\varepsilon/\mu} = (nm/\varepsilon)^4$ and very large. This means $Ax \leq (1+\varepsilon)\mathbb{1}$ must be true if the objective $f_\mu(x)$ is small.

We wish to point out that this is very different from the softmax function implicitly used in [You01], and is used as a potential function in [AK08a]. More precisely, the standard softmax function can be seen to arise as the Legendre dual of the negative entropy over the simplex, while our potential function is actually the Legendre dual of the negative *generalized* entropy over the positive quadrant. Our specific choice of this objective enables us to deduce what we call the multiplicative Lipschitz gradient property, described in (3.3).

We begin with several simple but important properties about OPT and $f_\mu(x)$. In short, they together imply that the minimum of $f_\mu(x)$ is around $-\text{OPT}$, and if one can approximately find the minimum of $f_\mu(x)$ (up to an error $O(\varepsilon\text{OPT})$), this corresponds to a $(1-O(\varepsilon))$-approximate solution to the packing LP (1.1). Notice that we will not be able to directly obtain a covering solution from this objective, and thus more techniques will be introduced in Section 4.

**Proposition 2.2.**
 (a) $\text{OPT} \in [1, n]$.
 (b) Letting $x = (1 - \varepsilon/2)x^* \geq 0$, we have $f_\mu(x) \leq -(1 - \varepsilon)\text{OPT}$.
 (c) Letting $x^{(0)} \geq 0$ be such that $x_i^{(0)} = \frac{1-\varepsilon/2}{n\|A_{:i}\|_\infty}$ for each $i \in [n]$, we have $f_\mu(x^{(0)}) \leq -\frac{1-\varepsilon}{n}$.
 (d) For any $x \geq 0$ satisfying $f_\mu(x) \leq 0$, we must have $Ax \leq (1+\varepsilon)\mathbb{1}$, and thus $\mathbb{1}^T x \leq (1+\varepsilon)\text{OPT}$.
 (e) If $x \geq 0$ satisfies $f_\mu(x) \leq -(1 - O(\varepsilon))\text{OPT}$, then $\frac{1}{1+\varepsilon}x$ is a $(1 - O(\varepsilon))$-approximate solution to the packing LP.
 (f) The gradient of $f_\mu(x)$ can be written as

$$\nabla f_\mu(x) = A^T y(x) - \mathbb{1} \quad \text{where} \quad y_j(x) \stackrel{\text{def}}{=} \exp^{\frac{1}{\mu}((Ax)_j - 1)} \quad . \tag{2.2}$$

(The proofs are straightforward and can be found in Appendix C.)

## 3 Parallelizable Packing LP Solver

In this section we prove the approximation and convergence guarantee on our packing LP algorithm. Although the same algorithm also produces a good covering LP solution, we defer such analysis to Section 4 because different techniques are required.

To describe our algorithm we first make the following choice of thresholding function

**Definition 3.1.** *The thresholding function* $\mathbb{T}\colon [-1, \infty) \to [-1, 1]$ *is defined as follows*

$$\mathbb{T}(v) \stackrel{\text{def}}{=} \begin{cases} 0, & v \in [-\varepsilon, \varepsilon]; \\ v, & v \in [-1, 1] \setminus [-\varepsilon, \varepsilon]; \\ 1, & v > 1. \end{cases}$$

Our algorithm is presented in Algorithm 1, and each of its iterations can be described with $x_i^{(k+1)} \leftarrow x_i^{(k)} \cdot \exp^{-\alpha \cdot \mathbb{T}(v_i)}$, where we choose $\alpha = \varepsilon\mu/4$ to be the step length. (Throughout this paper, we use superscript $x^{(k)}$ to represent vector $x$ at iteration $k$, and subscript $x_i$ to represent the $i$-th coordinate of vector $x$.)

Our proof of the correctness of `PosLPSolver` is divided into three steps.

**Step I: Gradient Descent.** We interpret (see Section 3.1 for details) each update $x_i^{(k+1)} \leftarrow x_i^{(k)} \cdot \exp^{-\alpha \cdot \mathbb{T}(v_i)}$ as a gradient descent step,[8] and show that the objective $f_\mu(x)$ does not increase,

---
[8]To be clear, in some literature, the gradient descent is referred only to $x \leftarrow x - c \cdot \nabla f(x)$ for some constant $c$. In this paper, we adopt the more general notion, and refer it to any step that directly decreases $f(x)$.

**Algorithm 1** `PosLPSolver`$(A, \varepsilon)$

**Input:** $A \in \mathbb{R}_{\geq 0}^{m \times n}, \varepsilon \in (0, 1/10]$.
**Output:** $x \in \mathbb{R}_{\geq 0}^n$ and $\overline{y} \in \mathbb{R}_{\geq 0}^m$.

1: $\mu \leftarrow \frac{\varepsilon}{4 \log(nm/\varepsilon)}$ and $\alpha \leftarrow \frac{\varepsilon \mu}{4}$. ⋄ *parameters*
2: $x_i^{(0)} \leftarrow \frac{1-\varepsilon/2}{n\|A_{:i}\|_\infty}$ for all $i \in [n]$. ⋄ *initial vector* $x^{(0)}$
3: $T \leftarrow \frac{6 \log(2n)}{\alpha \varepsilon}$. ⋄ *number of iterations*
4: **for** $k \leftarrow 0$ **to** $T - 1$ **do**
5:    **for** $i \leftarrow 1$ **to** $n$ **do**
6:       Compute the feedback $v_i \leftarrow \sum_{j=1}^m A_{i,j} \cdot \exp^{\frac{1}{\mu}((Ax)_j - 1)} - 1$
                                 ⋄ *in fact, $v_i = \nabla_i f_\mu(x^{(k)}) = \langle A_{:i}, y(x^{(k)}) \rangle - 1 \in [-1, \infty)$.*
7:       Update: $x_i^{(k+1)} \leftarrow x_i^{(k)} \cdot \exp^{-\alpha \cdot \mathbb{T}(v_i)}$. ⋄ *see Definition 3.1 for the definition of $\mathbb{T}(v)$*
8:    **end for**
9: **end for**
10: **return** $\frac{x^{(T)}}{1+\varepsilon}$ and $\overline{y} = \sum_{i=0}^{T-1} y(x^{(k)})$. ⋄ *recall that $y_j(x) \stackrel{\text{def}}{=} \exp^{\frac{1}{\mu}((Ax)_j - 1)}$*

or more strongly, always decreases by at least the following amount:

**Lemma 3.2** (Gradient Descent). *For any step $k$ in `PosLPSolver`, letting $B^{(k)} \subseteq [n]$ be the set of indices $i$ such that $\nabla_i f_\mu(x^{(k)}) \geq 1$, the objective $f_\mu(x)$ decreases by at least*
$$f_\mu(x^{(k)}) - f_\mu(x^{(k+1)}) \geq \frac{\alpha}{4} \cdot \sum_{i \in B^{(k)}} x_i^{(k)} \cdot \nabla_i f_\mu(x^{(k)}) \geq 0 \ .$$
*Combining this with Proposition 2.2.c, we have $f_\mu(x^{(k)}) \leq 0$ for all $k$.*

Note that the above gradient descent lemma does *not* follow from any classical theory because our objective $f_\mu(x)$ does not satisfy any good Lipschitz gradient property. Instead, we define and use a *multiplicative Lipschitz gradient property* for our objective, which may be of independent interest.

**Step II: Mirror Descent.** We interpret (see Section 3.2 for details) each update $x_i^{(k+1)} \leftarrow x_i^{(k)} \cdot \exp^{-\alpha \cdot \mathbb{T}(v_i)}$ as a mirror descent step.

A *mirror descent step* in optimization is any step from $x$ to $x'$ that is of the form $x' \leftarrow \arg\min_z \{V_x(z) + \langle \alpha \nabla f(x), z - x \rangle\}$. Here, $\alpha > 0$ is some step length, and $V_x(\widetilde{x}) = w(\widetilde{x}) - \langle \nabla w(x), \widetilde{x} - x \rangle - w(x)$ is the Bregman divergence of some convex *distance generating function* $w(x)$.[9] In this paper, we pick $w(x) \stackrel{\text{def}}{=} \sum_{i \in [n]} x_i \log x_i - x_i$ to be the generalized entropy function, and accordingly, for every $x, \widetilde{x} \geq 0$, let
$$V_x(\widetilde{x}) = \sum_{i \in [n]} \left( \widetilde{x}_i \log \frac{\widetilde{x}_i}{x_i} + x_i - \widetilde{x}_i \right) \ .$$
After verifying that our update is a mirror descent step, the next lemma easily follows from the general theory of mirror descent.

**Lemma 3.3** (Mirror Descent). *Letting $\xi_i^{(k)} \stackrel{\text{def}}{=} \mathbb{T}(\nabla_i f_\mu(x^{(k)})) \in [-1, 1]$ be the* truncated gradient, *we have that for any $u \geq 0$,*
$$\langle \alpha \xi^{(k)}, x^{(k)} - u \rangle \leq \alpha^2 \mathsf{OPT} + V_{x^{(k)}}(u) - V_{x^{(k+1)}}(u) \ .$$

We emphasize here that it is important to use the truncated gradient $\xi^{(k)} \in [-1, 1]^n$ in the mirror descent instead of the full gradient $\nabla f_\mu(x^{(k)})$, because the latter may have very large coordinates (whose magnitudes depend on the *width* of the matrix). This is why all previous positive-LP solvers using mirror descent are width-dependent. Our *gradient truncation* technique may be of independent interest.

---

[9] This $w(x)$ is classically chosen to be any *strongly convex* function, such as $w(x) = \frac{1}{2}\|x\|_2^2$ (and in that case $V_x(y) = \frac{1}{2}\|x - y\|_2^2$).

**Step III: Coupling.** Finally, as argued in Section 3.3, we put together the two lemmas above and derive the following coupled bound:

> **Lemma 3.4** (Coupling). *For any $u \geq 0$, we have*
> $$\alpha(f_\mu(x^{(k)}) - f_\mu(u)) \leq \langle \alpha \nabla f_\mu(x^{(k)}), x^{(k)} - u \rangle$$
> $$\leq 4(f_\mu(x^{(k)}) - f_\mu(x^{(k+1)})) + \big(V_{x^{(k)}}(u) - V_{x^{(k+1)}}(u)\big) + \alpha \cdot 2\varepsilon\mathsf{OPT} + \alpha \cdot \varepsilon \mathbb{1}^T u \ .$$

Let us point out right away that Lemma 3.4 captures benefit of combining the two analyses. If $f_\mu(x^{(k)}) - f_\mu(x^{(k+1)})$ is large, we are making a large gradient descent step because the objective greatly decreases. Or, if $f_\mu(x^{(k)}) - f_\mu(x^{(k+1)})$ is small (for a number of consecutive iterations), we can telescoping the above inequality and obtain a good upperbound on the average of $f_\mu(x^{(k)})$.

We are now ready to state and prove our theorem for packing LP.

> **Theorem 3.5** (Packing LP). *For $T \geq \frac{6\log(2n)}{\alpha\varepsilon} = \Omega(\frac{\log n \cdot \log(nm/\varepsilon)}{\varepsilon^3})$, we have that $f_\mu(x^{(T)}) \leq -(1-5\varepsilon)\mathsf{OPT}$, and as a consequence, $\mathtt{PosLPSolver}(A, \varepsilon)$ produces an output $x = \frac{x^{(T)}}{1+\varepsilon}$ that is a $(1 - O(\varepsilon))$-approximate solution for the packing LP (1.1).*

*Proof.* We begin by telescoping the inequality in Lemma 3.4 for $k = 0, 1, \ldots, T - 1$, and choosing $u = \widetilde{u} \stackrel{\text{def}}{=} (1 - \varepsilon/2)x^*$, which satisfies $\mathbb{1}^T u \leq \mathsf{OPT}$ by the definition of $x^*$:

$$\alpha \sum_{k=0}^{T-1} (f_\mu(x^{(k)}) - f_\mu(\widetilde{u})) \leq 4(f_\mu(x^{(0)}) - f_\mu(x^{(T)})) + \big(V_{x^{(0)}}(\widetilde{u}) - V_{x^{(T)}}(\widetilde{u})\big) + \alpha T \cdot 3\varepsilon\mathsf{OPT} \ . \quad (3.1)$$

Notice that, the second term on the right hand side is upper bounded by

$$V_{x^{(0)}}(\widetilde{u}) - V_{x^{(T)}}(\widetilde{u}) \leq V_{x^{(0)}}(\widetilde{u}) \leq \sum_i \widetilde{u}_i \log \frac{\widetilde{u}_i}{x_i^{(0)}} + x_i^{(0)} \leq \sum_i \widetilde{u}_i \log \frac{1/\|A_{:i}\|_\infty}{(1-\varepsilon/2)/n\|A_{:i}\|_\infty} + \frac{1-\varepsilon/2}{n\|A_{:i}\|_\infty}$$
$$\leq \mathbb{1}^T \widetilde{u} \cdot \log(2n) + 1 \leq 2\mathsf{OPT} \cdot \log(2n) \ . \quad (3.2)$$

Here, we have used the fact that $\widetilde{u}_i \leq \frac{1}{\|A_{:i}\|_\infty}$ since $A\widetilde{u} \leq \mathbb{1}$.

From here, we want to prove that $f_\mu(x^{(T)}) \leq -(1 - 5\varepsilon)\mathsf{OPT}$ by way of contradiction. Suppose not, that is, $f_\mu(x^{(T)}) > -(1 - 5\varepsilon)\mathsf{OPT}$, we have $f_\mu(x^{(0)}) - f_\mu(x^{(T)}) \leq 0 + (1 - 5\varepsilon)\mathsf{OPT} \leq \mathsf{OPT}$, giving an upper bound on the first term on the right hand side in (3.1). Substituting this and (3.2) to (3.1), and dividing $\alpha T$ on both sides, we get

$$\frac{1}{T} \sum_{k=0}^{T-1} (f_\mu(x^{(k)}) - f_\mu(\widetilde{u})) \leq \frac{4}{\alpha T}(f_\mu(x^{(0)}) - f_\mu(x^{(T)})) + \frac{1}{\alpha T}\big(V_{x^{(0)}}(\widetilde{u}) - V_{x^{(T)}}(\widetilde{u})\big) + 3\varepsilon\mathsf{OPT}$$
$$\leq \frac{4\mathsf{OPT}}{\alpha T} + \frac{2\mathsf{OPT} \cdot \log(2n)}{\alpha T} + 3\varepsilon\mathsf{OPT} \ .$$

Finally, since we have chosen $T \geq \frac{6\log(2n)}{\alpha\varepsilon}$, the above right hand side is no greater than $4\varepsilon\mathsf{OPT}$. This, by an averaging argument, tells us the existence of some $k \in \{0, 1, \ldots, T - 1\}$ with $f_\mu(x^{(k)}) \leq f_\mu(\widetilde{u}) + 4\varepsilon\mathsf{OPT} \leq -(1 - 5\varepsilon)\mathsf{OPT}$ (where we have used $f_\mu(\widetilde{u}) \leq -(1-\varepsilon)\mathsf{OPT}$ from Proposition 2.2.b). However, it contradicts to the hypothesis that $f_\mu(x^{(T)}) > -(1-5\varepsilon)\mathsf{OPT}$ because $f_\mu(x^{(k)}) \geq f_\mu(x^{(T)})$ according to Lemma 3.2. This finishes the proof that $f_\mu(x^{(T)}) \leq -(1 - 5\varepsilon)\mathsf{OPT}$. The fact that $\frac{x^{(T)}}{1+\varepsilon}$ provides a $(1 - O(\varepsilon))$ approximate solution for the packing LP is due to Proposition 2.2.e. □

## 3.1 The Gradient Descent Lemma

In this section, we are going to view our step $x^{(k)} \to x^{(k+1)}$ as a gradient descent step, and prove Lemma 3.2.

**Sketched Proof.** Here, we adopt a generalized notion of *gradient descent step*, and say that any step from $x$ to $x'$ that decreases the objective is a gradient descent step. Classically in optimization, if a convex function $f(x)$ satisfies the so-called Lipschitz gradient property, that is, $\|\nabla f(x_1) - \nabla f(x_2)\|_* \le L \cdot \|x_1 - x_2\|$ for some constant $L$ (with respect to some norm $\|\cdot\|$ and its dual norm $\|\cdot\|_*$), then a gradient descent step can provably decrease the objective by a considerable amount. (We refer interested readers to our survey in [AO17].) Unfortunately, this property is not obeyed by our objective $f_\mu(x)$, so we make use of what we call the *multiplicative Lipschitz gradient* property, that may be of independent interest for convex optimization problems that have enough 'non-negativity'.

In particular, we observe that:

> In each iteration, `PosLPSolver` changes each coordinate of $x$ *multiplicatively* by at most a factor of $1 \pm 4\alpha/3$. Owing to our choice of the smoothed objective $f_\mu(x)$, we can prove that in this iteration, for each $i$ satisfying $|\nabla_i f_\mu(x)| > \varepsilon$, the coordinate gradient
>
> $$\nabla_i f_\mu(x) \text{ is not changed by more than a multiplicative factor of } 1 \pm 0.5. \tag{3.3}$$

Denoting by $x = x^{(k)}$ the vector before the update, and $x' = x^{(k+1)}$ the one after, let us now estimate the difference between $f_\mu(x) - f_\mu(x')$ using (3.3), and sketch the proof of Lemma 3.2.

Since $\nabla f_\mu(x)$ is close enough to $\nabla f_\mu(x')$ owing to (3.3), intuitively, we can show that $f_\mu(x) - f_\mu(x')$ is (up to a constant factor) close to $\langle \nabla f_\mu(x), x - x' \rangle$ due to the first-order approximation of $f_\mu(x)$ around $x$. Now, since $x_i - x'_i$ is positive only when $\nabla_i f_\mu(x)$ is positive, and viceversa, we conclude that the difference $f_\mu(x) - f_\mu(x') \approx \langle \nabla f_\mu(x), x - x' \rangle$ is non-negative.

Furthermore, when focusing only on the coordinates $i$ such that $\nabla_i f_\mu(x) \ge 1$ (i.e., $i \in B^{(k)}$), we have that $x_i - x'_i = x_i(1 - e^{-\alpha}) = \Omega(\alpha) \cdot x_i$. This enables us to conclude that the amount of difference $f_\mu(x) - f_\mu(x')$ is at least $\Omega(\alpha) \cdot \sum_{i \in B_{(k)}} x_i \cdot \nabla_i \mu(x)$, arriving at the conclusion of Lemma 3.2.

**Proof Details.** The following proposition establishes the formal statement for (3.3).

**Proposition 3.6.** *If $f_\mu(x^{(k)}) \le 0$, for any $x = \tau x^{(k)} + (1-\tau)x^{(k+1)}$ where $\tau \in [0,1]$:*

(a) $x_i \in x_i^{(k)} \cdot [1 - 4\alpha/3, 1 + 4\alpha/3]$

(b) $y_j(x) \in y_j(x^{(k)}) \cdot [1 - \varepsilon/2, 1 + \varepsilon/2]$

(c) *When* $|\nabla_i f_\mu(x^{(k)})| \ge \varepsilon$, *we have that* $\nabla_i f_\mu(x)$ *is between* $\frac{\nabla_i f_\mu(x^{(k)})}{2}$ *and* $\frac{3\nabla_i f_\mu(x^{(k)})}{2}$.

*Proof.*

(a) We can always write $x_i = x_i^{(k)} \cdot e^t$ for some $t \in [-\alpha, \alpha] \subseteq [-1/4, 1/4]$. According to the fact that $e^t \le 1 + 4t/3$ for $t \in [0, 1/4]$ and $e^t \ge 1 - t \ge 1 - 4t/3$ for $t \in [-1/4, 0]$, we must have $x_i \in x_i^{(k)} \cdot [1 - 4\alpha/3, 1 + 4\alpha/3]$.

(b) Recall from (2.2) that $y_j(x) = \exp^{\frac{1}{\mu}((Ax)_j - 1)}$. According to Proposition 2.2.d, we have $(Ax^{(k)})_j \le 1 + \varepsilon$. Now, by the non-negativity of $A$ and the previous item, we have

$$\left|(Ax)_j - (Ax^{(k)})_j\right| \le 4\alpha/3 \cdot (Ax^{(k)})_j \le 4\alpha/3 \cdot (1 + \varepsilon) \le 5\alpha/3 \ .$$

This implies that $y_j(x) \ge y_j(x^{(k)}) \cdot \exp(-5\alpha/3\mu) = y_j(x^{(k)}) \cdot \exp(-5\varepsilon/12) > y_j(x^{(k)}) \cdot (1 - \varepsilon/2)$ for sufficiently small $\varepsilon$, as well as that $y_j(x) \le y_j(x^{(k)}) \cdot \exp(5\alpha/3\mu) < y_j(x^{(k)}) \cdot (1 + \varepsilon/2)$.

(c) Recall from (2.2) that $\nabla_i f_\mu(x) = (A^T y(x))_i - 1$. By symmetry, we only prove the case when $\nabla_i f_\mu(x^{(k)}) \geq \varepsilon$, which is equivalent to $(A^T y(x^{(k)}))_i \geq 1 + \varepsilon$. By the previous item, we immediately have

$$(A^T y(x^{(k)}))_i (1 + \varepsilon/2) \geq (A^T y(x))_i \geq (A^T y(x^{(k)}))_i (1 - \varepsilon/2) \ .$$

Denoting by $t = (A^T y(x^{(k)}))_i - 1 \geq \varepsilon$, it is not hard to verify that $(t+1)(1-\varepsilon/2) \geq t/2 + 1$ and $(t+1)(1+\varepsilon/2) \leq 3t/2 + 1$ for all $t \geq \varepsilon$, which then implies

$$\frac{3\nabla_i f_\mu(x^{(k)})}{2} = 3t/2 \geq (A^T y(x))_i - 1 \geq t/2 = \frac{\nabla_i f_\mu(x^{(k)})}{2} \qquad \square$$

We can now use the above multiplicative Lipschitz gradient property to prove the desired gradient descent progress promised in Lemma 3.2.

*Proof of Lemma 3.2.* We prove by induction. Suppose that Lemma 3.2 is true for all indices less than $k$. This implies, in particular, that $f_\mu(x^{(k)}) \leq f_\mu(x^{(k-1)}) \leq \cdots \leq f_\mu(x^{(0)}) \leq 0$.

We compute the objective difference by the standard integral over gradients as follows.

$$f_\mu(x^{(k)}) - f_\mu(x^{(k+1)}) = \int_0^1 \left\langle \nabla f_\mu(x^{(k+1)} + \tau(x^{(k)} - x^{(k+1)})), x^{(k)} - x^{(k+1)} \right\rangle d\tau$$

$$= \sum_i \int_0^1 \nabla_i f_\mu(x^{(k+1)} + \tau(x^{(k)} - x^{(k+1)})) d\tau \times (x_i^{(k)} - x_i^{(k+1)}) \geq 0 \qquad (3.4)$$

Here the last inequality is because, whenever $x_i^{(k)} - x_i^{(k+1)}$ is strictly positive (resp. strictly negative) for some coordinate $i \in [n]$, it must be because $\nabla_i f_\mu(x^{(k)}) \geq \varepsilon$ (resp. $\leq -\varepsilon$) according to our algorithm. However, owing to Proposition 3.6.c, we have that $f_\mu(x^{(k+1)} + \tau(x^{(k)} - x^{(k+1)}))$ is also positive (resp. negative) for all $\tau \in [0, 1]$, since $\nabla_i f_\mu(x^{(k)})$ is. (Here we used $f_\mu(x^{(k)}) \leq 0$.) This concludes that for each $i$, the $i$-th component in (3.4), denoted by $W_i \stackrel{\text{def}}{=} \int_0^1 \nabla_i f_\mu(x^{(k+1)} + \tau(x^{(k)} - x^{(k+1)})) d\tau \times (x_i^{(k)} - x_i^{(k+1)})$, is non-negative.

We next turn to lower bounding $f_\mu(x^{(k)}) - f_\mu(x^{(k+1)})$ by computing a lower bound on $W_i$ for each $i \in B^{(k)}$. Indeed, recall that by the definition of our thresholding function $\mathbb{T}(\cdot)$, for each $i \in B^{(k)}$, the update on the $i$-th coordinate in $x^{(k)}$ is precisely $x_i^{(k+1)} \leftarrow x_i^{(k)} \cdot \exp^{-\alpha}$. In such a case,

$$W_i = (1 - e^{-\alpha}) x_i^{(k)} \times \int_0^1 \nabla_i f_\mu(x^{(k+1)} + \tau(x^{(k)} - x^{(k+1)})) d\tau$$

$$\geq (1 - e^{-\alpha}) x_i^{(k)} \times \frac{1}{2} \nabla_i f_\mu(x^{(k)}) \qquad \text{(using Proposition 3.6.c)}$$

$$\geq \frac{\alpha}{4} \cdot x_i^{(k)} \cdot \nabla_i f_\mu(x^{(k)}) \ .$$

In sum, we conclude that

$$\sum_i W_i \geq \frac{\alpha}{4} \cdot \sum_{i \in B^{(k)}} x_i^{(k)} \cdot \nabla_i f_\mu(x^{(k)}) \ . \qquad \square$$

### 3.2 The Mirror Descent Lemma

In this section, we are going to view our step $x^{(k)} \to x^{(k+1)}$ as a mirror descent step, and prove Lemma 3.3.

Recall that $\xi_i^{(k)} \stackrel{\text{def}}{=} \mathbb{T}(\nabla_i f_\mu(x^{(k)})) \in [-1, 1]$ is the truncated gradient at step $k$, and satisfies that $\xi_i^{(k)} = \nabla_i f_\mu(x^{(k)})$ for all coordinates $i$ such that $\nabla_i f_\mu(x^{(k)}) \in [-1, 1] \setminus [-\varepsilon, \varepsilon]$. We can verify that our careful choice of $x^{(k)} \to x^{(k+1)}$ is in fact a mirror descent step on the truncated gradient:

**Claim 3.7.**
$$x^{(k+1)} = \arg\min_{z \geq 0} \left\{ V_{x^{(k)}}(z) + \langle \alpha \xi^{(k)}, z - x^{(k)} \rangle \right\} . \tag{3.5}$$

*Proof.* This can be verified coordinate by coordinate, because the arg min function is over all possible $z \geq 0$, where this constraint does not impose any inter-coordinate constraint.

In other words, by substituting the definition of $V_{x^{(k)}}(z)$, we only need to verify that
$$x_i^{(k+1)} = \arg\min_{z_i \geq 0} \left\{ \left( z_i \log \frac{z_i}{x_i^{(k)}} + x_i^{(k)} - z_i \right) + \alpha \xi_i^{(k)} \cdot (z_i - x_i^{(k)}) \right\} \stackrel{\text{def}}{=} \arg\min_{z_i \geq 0} \{g(z_i)\} .$$
At this point, the univariate function $g(z_i)$ is convex and has a unique minimizer. Since the gradient $\frac{d}{dz_i} g(z_i) = \log \frac{z_i}{x_i^{(k)}} + \alpha \xi_i^{(k)}$, this unique minimizer is indeed $z_i = x_i^{(k)} \cdot \exp^{-\alpha \xi_i^{(k)}}$, finishing the proof of Claim 3.7. □

After confirming that our iterative step in `PosLPSolver` is indeed a mirror descent step, it is not hard to deduce Lemma 3.3 based on the proof of the classical mirror descent analysis (see for instance [BN13]). However, we emphasize here that our choice of the distance generating function $w(x)$ is not strongly convex over the entire positive orthant $\{x \in \mathbb{R}^n : x \geq 0\}$, and thus the our proof is not identical to the classical theory. We have relied on, in fact, a 'local' strong convexity which we introduce and is sufficient for our purpose (see (3.6)).

*Proof of Lemma 3.3.* We deduce the following sequence of inequalities:
$$\langle \alpha \xi^{(k)}, x^{(k)} - u \rangle = \langle \alpha \xi^{(k)}, x^{(k)} - x^{(k+1)} \rangle + \langle \alpha \xi^{(k)}, x^{(k+1)} - u \rangle$$
$$\stackrel{①}{=} \langle \alpha \xi^{(k)}, x^{(k)} - x^{(k+1)} \rangle + \langle -\nabla V_{x^{(k)}}(x^{(k+1)}), x^{(k+1)} - u \rangle$$
$$\stackrel{②}{=} \langle \alpha \xi^{(k)}, x^{(k)} - x^{(k+1)} \rangle + V_{x^{(k)}}(u) - V_{x^{(k+1)}}(u) - V_{x^{(k)}}(x^{(k+1)})$$
$$\stackrel{③}{\leq} \sum_i \left( \alpha \xi_i^{(k)} \cdot (x^{(k)} - x^{(k+1)}) - \frac{|x_i^{(k+1)} - x_i^{(k)}|^2}{2 \max\{x_i^{(k+1)}, x_i^{(k)}\}} \right) + \left( V_{x^{(k)}}(u) - V_{x^{(k+1)}}(u) \right)$$
$$\stackrel{④}{\leq} \sum_i \frac{(\alpha^2 \xi_i^{(k)})^2 \cdot \max\{x_i^{(k+1)}, x_i^{(k)}\}}{2} + \left( V_{x^{(k)}}(u) - V_{x^{(k+1)}}(u) \right) \tag{3.6}$$
$$\stackrel{⑤}{\leq} \frac{2}{3} \alpha^2 \mathbb{1}^T x^{(k)} + \left( V_{x^{(k)}}(u) - V_{x^{(k+1)}}(u) \right)$$
$$\stackrel{⑥}{\leq} \alpha^2 \mathsf{OPT} + \left( V_{x^{(k)}}(u) - V_{x^{(k+1)}}(u) \right)$$

Here, ① is due to the minimality of $x^{(k+1)}$ in (3.5), which implies that $\nabla V_{x^{(k)}}(x^{(k+1)}) + \alpha \xi^{(k)} = 0$. ② is due to the triangle equality of Bregman divergence:
$$\forall x, y \geq 0, \quad \langle -\nabla V_x(y), y - u \rangle = \langle \nabla w(x) - \nabla w(y), y - u \rangle$$
$$= (w(u) - w(x) - \langle \nabla w(x), u - x \rangle) - (w(u) - w(y) - \langle \nabla w(y), u - y \rangle)$$
$$- (w(y) - w(x) - \langle \nabla w(x), y - x \rangle)$$
$$= V_x(u) - V_y(u) - V_x(y) .$$
③ is because $V_x(y) = \sum_i y_i \log \frac{y_i}{x_i} + x_i - y_i \geq \sum_i \frac{1}{2 \max\{x_i, y_i\}} |x_i - y_i|^2$. ④ is by Cauchy-Schwarz. ⑤ is because we have $x_i^{(k+1)} \leq \frac{4}{3} x_i^{(k)}$ owing to Proposition 3.6.a. ⑥ is because we have $\mathbb{1}^T x^{(k)} \leq \frac{3}{2} \mathsf{OPT}$ owing to Proposition 2.2.d (and $f_\mu(x^{(k)}) \leq 0$ from Lemma 3.3). □

**Remark 3.8.** The main difference between this proof and its classical counterpart in optimization theory is inequality ③ in (3.6). Recall that $w(x) = \sum_{i=1}^{n} x_i \log x_i - x_i$. Since $w(x)$ is known to be 1-strongly convex with respect to the $\ell_1$-norm over the simplex $\Delta = \{x \geq 0 : \mathbb{1}^T x = 1\}$, we automatically have $V_x(y) \geq \frac{1}{2}\|x-y\|_1^2$ for all $x, y \in \Delta$, and this was the key step used in the classical analysis. In our case, we no longer have this strong convexity because $x, y \notin \Delta$. However, the fact that $V_x(y) \geq \sum_i \frac{1}{2\max\{x_i, y_i\}} |x_i - y_i|^2$ is in fact saying that $w(x)$ is 'locally' 1-strongly convex with respect to the $\|\cdot\|_{x,2}$ norm, defined to be $\|w\|_{x,2}^2 \stackrel{\text{def}}{=} \sum_i w_i^2/x_i$. This local norm technique is very crucial in our analysis, and is the optimization-based intuition behind the above lemma.

## 3.3 The Coupling Lemma

The main idea in our proof to Lemma 3.4 is to divide the gradient vector $\nabla f(x) \in [-1, \infty)^n$ into three components, the component containing large coordinates (i.e., bigger than 1), the component containing small coordinates (i.e., in $[-1, 1] \setminus [-\varepsilon, \varepsilon]$), and the component containing negligible coordinates (i.e., in $[-\varepsilon, \varepsilon]$). The large gradients are to be taken care by the gradient descent lemma, the small gradients are to be taken care by the mirror descent lemma. Formally,

*Proof of Lemma 3.4.* By convexity, the distance $f_\mu(x^{(k)}) - f_\mu(u)$ for an arbitrary $u \geq 0$ is upper bounded as follows:

$$\alpha(f_\mu(x^{(k)}) - f_\mu(u)) \leq \langle \alpha \nabla f_\mu(x^{(k)}), x^{(k)} - u \rangle$$
$$= \langle \alpha \eta^{(k)}, x^{(k)} - u \rangle + \langle \alpha \xi^{(k)}, x^{(k)} - u \rangle + \langle \alpha \zeta^{(k)}, x^{(k)} - u \rangle , \quad (3.7)$$

where

- $\xi_i^{(k)} \stackrel{\text{def}}{=} \mathbb{T}(\nabla_i f_\mu(x^{(k)})) \in [-1, 1]$ is the *truncated gradient*, capturing the small coordinates.
- $\eta_i^{(k)} \stackrel{\text{def}}{=} \left\{ \begin{array}{ll} \nabla_i f_\mu(x^{(k)}) - \xi_i^{(k)}, & \text{if } \nabla_i f_\mu(x^{(k)}) \geq 1; \\ 0, & \text{otherwise.} \end{array} \right\} \in [0, \infty)$, capturing the large coordinates.
- $\zeta_i^{(k)} \stackrel{\text{def}}{=} \nabla_i f_\mu(x^{(k)}) - \xi_i^{(k)} - \eta_i^{(k)} \in [-\varepsilon, \varepsilon]$, capturing the negligible coordinates.

We analyze the three components of (3.7) one by one.

The $\zeta$ component is small: if $f_\mu(u) \leq 0$, we have

$$\langle \alpha \zeta^{(k)}, x^{(k)} - u \rangle \leq \alpha \varepsilon \cdot (\mathbb{1}^T x^{(k)} + \mathbb{1}^T u) \leq \alpha \varepsilon \cdot (1 + \varepsilon)\mathsf{OPT} + \alpha \varepsilon \cdot \mathbb{1}^T u \quad (3.8)$$

where the last inequality is because $f_\mu(x^{(k)}) \leq 0$ from Lemma 3.2.

The $\eta$ component can be upper bounded with the help from our gradient descent Lemma 3.2. Note that $\eta_i^{(k)} \neq 0$ only if $i \in B^{(k)}$ (where recall from Lemma 3.2 that $B^{(k)}$ is the set of indices whose $\nabla_i f_\mu(x^{(k)})$ is no less than 1). In particular, if $i \in B^{(k)}$ we have $\eta_i^{(k)} = \nabla_i f_\mu(x^{(k)}) - 1 < \nabla_i f_\mu(x^{(k)})$, and thus Lemma 3.2 gives

$$\frac{4(f_\mu(x^{(k)}) - f_\mu(x^{(k+1)}))}{\alpha} \geq \sum_{i \in B^{(k)}} x_i^{(k)} \cdot \nabla_i f_\mu(x^{(k)}) \geq \langle \eta^{(k)}, x^{(k)} \rangle$$

$$\implies \langle \alpha \eta^{(k)}, x^{(k)} - u \rangle \leq \langle \alpha \eta^{(k)}, x^{(k)} \rangle \leq 4(f_\mu(x^{(k)}) - f_\mu(x^{(k+1)}))$$

Finally, the $\xi$ component is upper bounded by Lemma 3.3. Together, we obtain

$$\alpha(f_\mu(x^{(k)}) - f_\mu(u)) \leq \langle \alpha \eta^{(k)}, x^{(k)} - u \rangle + \langle \alpha \xi^{(k)}, x^{(k)} - u \rangle + \langle \alpha \zeta^{(k)}, x^{(k)} - u \rangle$$
$$\leq 4(f_\mu(x^{(k)}) - f_\mu(x^{(k+1)})) + \alpha^2 \mathsf{OPT} + V_{x^{(k)}}(u) - V_{x^{(k+1)}}(u) + \alpha \varepsilon \cdot (1+\varepsilon)\mathsf{OPT} + \alpha \varepsilon \mathbb{1}^T u$$
$$\leq 4(f_\mu(x^{(k)}) - f_\mu(x^{(k+1)})) + \left(V_{x^{(k)}}(u) - V_{x^{(k+1)}}(u)\right) + \alpha \cdot 2\varepsilon \mathsf{OPT} + \alpha \cdot \varepsilon \mathbb{1}^T u . \quad \square$$

## 4 Parallelizable Covering LP Solver

Since a primal solution $x$ satisfying $f_\mu(x) \approx -\mathsf{OPT}$ does not translate into a dual solution $y$ of the covering LP (1.2), the results in Section 3 do not imply any good approximate to the covering LP program. In fact, most of the previous results (except Luby and Nisan) have encountered this similar problem, and thus needed a separate algorithm to solve the covering LP. We show in this section that, in our same algorithm `PosLPSolver`, once the average $\overline{y} = \sum_{i=0}^{T-1} y(x^{(k)})$ is collected over all the iterations, this $\overline{y}$ is essentially an approximate solution to the covering LP.

The high level intuition behind this result is very clear. On one hand, the packing LP (1.1) is dual to the covering LP (1.2). On the other hand, `PosLPSolver` falls into a primal-dual framework: (a) the (primal) gradient descent ensures that the final objective $f_\mu(x^{(T)})$ is sufficiently small, while (b) the (dual) mirror descent ensures that the average of the encountered gradients (which is a function on $\overline{y}$) is sufficiently close to 0. If (a) gives rise to an approximate solution to the packing LP, then (b) should, at least intuitively, give rise to a dual solution $\overline{y}$ of the covering LP.

More formally, after telescoping Lemma 3.4 for $k = 0, 1, \ldots, T-1$, we have for any $u \geq 0$,

$$\frac{1}{T} \sum_{k=0}^{T-1} \langle \nabla f_\mu(x^{(k)}), x^{(k)} - u \rangle \leq \frac{4}{\alpha T}(f_\mu(x^{(0)}) - f_\mu(x^{(T)})) + \frac{1}{\alpha T}\big(V_{x^{(0)}}(u) - V_{x^{(T)}}(u)\big) + 2\varepsilon\mathsf{OPT} + \varepsilon\mathbb{1}^T u$$

$$\leq \frac{4}{\alpha T}(f_\mu(x^{(0)}) - f_\mu(x^{(T)})) + \frac{1}{\alpha T}V_{x^{(0)}}(u) + 2\varepsilon\mathsf{OPT} + \varepsilon\mathbb{1}^T u \ . \quad (4.1)$$

This upper bound (on the average regret) gives a lot of information about the average gradient $\frac{1}{T}\sum_k \nabla f_\mu(x^{(k)})$, thanks to the arbitrary choice of $u \geq 0$. For instance, if most of the terms in (4.1) were zero and we had $\frac{1}{T}\sum_{k=0}^{T-1}\langle \nabla f_\mu(x^{(k)}), -u\rangle \leq 0$, we would have $\frac{1}{T}\sum_k \nabla f_\mu(x^{(k)}) \geq 0$, which is equivalent to $A^T\overline{y} \geq \mathbb{1}$, the feasibility of the covering LP. However, since there are five missing terms in this wishful example, more careful studies are needed.

It is worth noting that the average $\overline{y}$ only provides a $(1 + O(\varepsilon))$ approximation to the covering LP when $T \geq \Omega(\frac{\log(n\rho)\log(nm/\varepsilon)}{\varepsilon^3})$, where $\rho$ is the width of $A$. This is slightly worse than the $T$ required in Algorithm 1, because $\log(n\rho)$ may in principle be slightly larger than $\log(n)$. We prove, however, if one is willing to perform a linear time coordinate fixing on the output $\overline{y}$, then the same number of iterations from Algorithm 1 is sufficient. This result requires a more careful choice of $u \geq 0$ in the above reasoning.

We defer all the technical details on the covering LP including the formal statement of our theorem (see Theorem D.3 on page 22) to Appendix D.

## Acknowledgements

We thank Jonathan Kelner, Yin Tat Lee, Richard Peng, and Neal Young for helpful conversations. This material is based upon work partly supported by the National Science Foundation under Grant CCF-1319460 and by a Simons Graduate Student Award under grant no. 284059.

# Appendix

## A  Empirical Evaluation

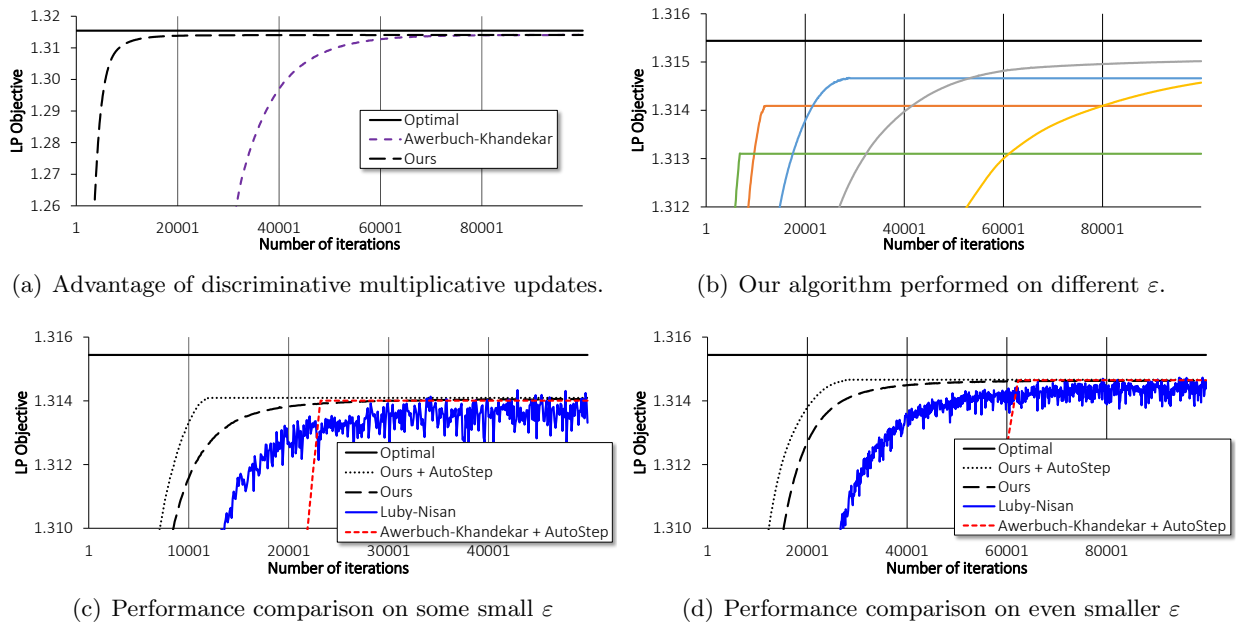

(a) Advantage of discriminative multiplicative updates.

(b) Our algorithm performed on different $\varepsilon$.

(c) Performance comparison on some small $\varepsilon$

(d) Performance comparison on even smaller $\varepsilon$

Figure 1: Empirical Evaluations

### A.1  AutoStep: Automatic Step-Length Computation

We begin this section by describing an implementation trick that can be applied to both our algorithm and Awerbuch and Khandekar [AK08a]. Recall from (3.3) that, we have chosen our $\alpha$ in `PosLPSolver` to be the (theoretically) most aggressive value such that $\nabla_i f_\mu(x)$ is not going to be affected multiplicatively by more than $1 \pm 0.5$. In practice, however, this maximal step length $\alpha$ can be computed numerically during each iteration, and can be made different among iterations.[10] This automatic step-length computation can also be applied to Awerbuch and Khandekar [AK08a], and has already been implicitly applied to all other previous algorithms.[11]

### A.2  Illustration

We perform some simple experiments to illustrate the performance of our new algorithm with real data. We focus on the packing LP program with a randomly generated matrix $A \in \mathbb{R}^{60 \times 40}$ of

---

[10]It is even true that our theorems can be adapted to allow different $\alpha$'s to be used, however, we have chosen not to do so for the simplicity of our theoretical results.

[11]Other algorithms —namely, [LN93, BBR04, You01]— have implemented this automatic step-length computation for a different purpose: they need it in their convergence analysis but we do not. This is one of the reasons our algorithm `PosLPSolver` is much simpler than theirs. (In their algorithms, the convergence analysis is quite combinatorial and works essentially as follows. In each iteration, because the update rule is maximally aggressive, at least one of the inner products $\langle A_i, x \rangle$ is going to be increased by a fixed additive amount. However, this increment cannot happen too many times because otherwise at least one of the constraints will be violated.)

800 non-zero entries each in the range of $[0, 10]$, whose optimal value $\mathsf{OPT} = 1.31544$. We have implemented the following five algorithms.

- Luby and Nisan [LN93].
- Awerbuch and Khandekar [AK08a], with and without the AutoStep trick.
- Our `PosLPSolver`, with and without the AutoStep trick.

**Importance of Discriminative Updates.** We compare the solver of Awerbuch and Khandekar with ours, to illustrate the importance of using *discriminative* multiplicative updates. (Recall that the algorithm of Awerbuch and Khandekar [AK08a] is very similar to ours, except that they update all the relevant coordinates by the same factor, while we treat them differently and update a coordinate $x_i$ more slowly if its feedback $v_i$ is small.) Figure 1(a) clearly confirms that this discrimination is very important.

**Role of the Smoothed Objective.** Notice that, for our algorithm `PosLPSolver`, when the input parameter $\varepsilon$ varies, the performance curves go *across* each other (see Figure 1(b)). To be clear, with larger $\varepsilon$ the curve goes up faster but converges to a worse solution (see the bottommost green curve); while on the other hand, with smaller $\varepsilon$ the curve goes up slower but has the potential to converge to a better solution (see the rightmost orange curve). This is because, for different values of $\varepsilon$, our smoothed objective $f_\mu(x)$ has its parameter $\mu$ dependent on $\varepsilon$, and therefore the minimum points of $f_\mu(x)$ will have different distances to the actual LP optimum.

(This behavior is in fact shared with all other methods as well.[12] Therefore, to conduct a fair experiment when comparing different algorithms in the next paragraph, we tune the input parameters —via binary search— on each algorithm separately, so as to make sure that they converge to the same value. Then, we plot the curves corresponding to these input parameters.)

**Performance Comparison.** We illustrate the performance difference between Luby-Nisan, our `PosLPSolver` (with and without AutoStep), and Awerbuch-Khandekar with AutoStep. We have ignored Awerbuch-Khandekar in this comparison due to its poor performance. We have chosen two quite small values of $\varepsilon$ in order to clearly see the performance difference between algorithms that have different dependencies on $\varepsilon$. It is clear from Figure 1(c) and Figure 1(d) that our algorithm outperforms all others, and the practical performance of AutoStep is also considerable. It is worth noting that the solution produced by `PosLPSolver` is much more stable than Luby-Nisan (because we focus on the decreasing of some objective $f_\mu(x)$ while their algorithm is quite combinatorial), and each iteration of ours is at least 5 times faster than theirs due to the simplicity of our algorithm.

## B  Semi-Stateless Feature of our Positive-LP Solver

One typical distributed setting for implementing a parallelizable positive-LP solver is as follows.[13] Suppose that there is an agent $i$ controlling variable $x_i$, and agent $i$ is assumed to know (1) (upper bounds on) $m$ and $n$, (2) the $i$-th column of $A$, and (3) the current "congestion" $(Ax)_j$ for those constraints $j$ that agent $i$ has non-zero influence (i.e., for those $j$ such that $A_{i,j} > 0$). These are the only information disclosed to agent $i$.

It is not hard to verify that our `PosLPSolver`$(A, \varepsilon)$, like most of the previous results in Table 1, can be implemented in this distributed setting in $\frac{\log^2 N}{\varepsilon^3}$ synchronized iterations.

---

[12] All known methods are implicitly 'smoothing' the LP objective by some parameter, and then performing the related updates. Therefore, none of our algorithms converge to the LP optimum.

[13] We refer interested readers to [AK08a] for the strong motivations and practical examples for such settings.

**Stateless Algorithms.** Recently, distributed algorithms that are *stateless* have received a lot of attention [AK08a, AAK08, AK08b, AK09]. In the language of positive LPs (see [AK08a]), the stateless requirement says that

> "the decisions made by agents are not dependent on the past;
> they are only dependent on the current local state observable to the agents."

Although their definition is vague, statelessness implies the following three important properties, and therefore to check if an algorithm is stateless, it suffices to verify them one by one.

(P1) *Self-stabilization.* The algorithm is robust against adversarial but finite sequence of "hard reset" events. This allows some agents to fall asleep for a finite period of time, and then to wake up; or equivalently, it means that the algorithm does not need to be initialized.

(P2) *Robustness against incremental adjustments.* Agents are allowed to join or leave dynamically. This corresponds to zeroing out or introducing new columns in $A$, without restarting other agents. Adding or deleting rows, or even modifications to entries of $A$ are similarly allowed.

(P3) *No global clock.* Algorithms can proceed asynchronously without a global clock.

Before Awerbuch and Khandekar [AK08a], all known parallelizable positive-LP solvers are stateful, and do not satisfy any of the three properties above. In particular, the width-independent ones are phaseful and have to inform each agent 'which phase it is in' (and many of them only increase $x$ throughout the process), while the width-dependent ones (such as [PST95]) must keep track of the maximum violation in a constraint.

**Our Semi-Stateless Positive-LP Solver.** We wish to point out that our `PosLPSolver` can be easily tuned to at least satisfy (P1) and (P2). However, our current analysis still requires the agents to act synchronously and therefore needs a global clock. We call any algorithm that satisfy (P1) and (P2) *semi-stateless*.[14]

Indeed, the only line we need to change in the algorithm `PosLPSolver`$(A, \varepsilon)$ is to let

$$x_i^{(k+1)} \leftarrow \max\left\{x_i^{(k)} \cdot \exp^{-\alpha \cdot \mathbb{T}(v_i)}, \frac{\delta}{\|A_{:i}\|_\infty}\right\} ,$$

where $\delta$ is some small enough number such as $\delta = (\varepsilon/nm)^5$. This small modification was also used in [AK08a] to obtain stateless algorithms, and makes our algorithm robust again arbitrarily chosen input. (For instance, adversarially chosen agents may initialize some coordinate $x_i$ to zero; without the introduction of $\delta$, the value of $x_i$ will freeze at zero since each step is only multiplicative.)

We ignore the formal proof of statelessness in this version of the paper because it is routinary.

## C Missing Proof of Proposition 2.2

**Proposition 2.2.**
 (a) $\mathsf{OPT} \in [1, n]$.
 (b) Letting $x = (1 - \varepsilon/2)x^* \geq 0$, we have $f_\mu(x) \leq -(1-\varepsilon)\mathsf{OPT}$.
 (c) Letting $x^{(0)} \geq 0$ be such that $x_i^{(0)} = \frac{1-\varepsilon/2}{n\|A_{:i}\|_\infty}$ for each $i \in [n]$, we have $f_\mu(x^{(0)}) \leq -\frac{1-\varepsilon}{n}$.
 (d) For any $x \geq 0$ satisfying $f_\mu(x) \leq 0$, we must have $Ax \leq (1+\varepsilon)\mathbb{1}$, and thus $\mathbb{1}^T x \leq (1+\varepsilon)\mathsf{OPT}$.

---

[14]Technically speaking, the agents in our algorithm `PosLPSolver` do not have states as well, but do need to use a virtual global state that is the clock.

(e) If $x \geq 0$ satisfies $f_\mu(x) \leq -(1 - O(\varepsilon))\mathsf{OPT}$, then $\frac{1}{1+\varepsilon}x$ is a $(1 - O(\varepsilon))$-approximate solution for the packing LP.

(f) The gradient of $f_\mu(x)$ can be written as
$$\nabla f_\mu(x) = A^T y(x) - \mathbb{1} \quad \text{where} \quad y_j(x) \stackrel{\text{def}}{=} \exp^{\frac{1}{\mu}((Ax)_j - 1)} \enspace.$$

*Proof.*

(a) Suppose that $i^*$ is the column that achieves the smallest infinite norm $\|A_{:i}\|_\infty$ over all columns. Letting $x$ be such that $x_i = 1$ at $i = i^*$ and $x_i = 0$ elsewhere, we have obtained a feasible solution for the packing LP (1.1), owing to our choice of $\min_{i \in [n]}\{\|A_{:i}\|_\infty\} = 1$ in (2.1). This feasible $x$ gives an objective $\mathbb{1}^T x = 1$, showing that $\mathsf{OPT} \geq 1$.

On the other hand, for any solution $x \in \mathbb{R}^n_{\geq 0}$ satisfying $Ax \leq \mathbb{1}$, we must have $x_i \leq \frac{1}{\|A_{:i}\|_\infty}$ for each $i$. Therefore, $\mathbb{1}^T x \leq \sum_i \frac{1}{\|A_{:i}\|_\infty} \leq n$, showing that $\mathsf{OPT} \leq n$.

(b) We have $\mathbb{1}^T x = (1 - \varepsilon/2)\mathsf{OPT}$ by the definition of $\mathsf{OPT}$. Also, from the feasibility constraint $Ax^* \leq \mathbb{1}$ in the packing LP, we have $Ax - \mathbb{1} \leq -\varepsilon/2 \cdot \mathbb{1}$, and can compute $f_\mu(x)$ as follows:
$$f_\mu(x) = \mu \sum_j \exp^{\frac{1}{\mu}((Ax)_j - 1)} - \mathbb{1}^T x \leq \mu \sum_j \exp^{\frac{-\varepsilon/2}{\mu}} - (1 - \varepsilon/2)\mathsf{OPT}$$
$$\leq \frac{\mu m}{(nm)^2} - (1 - \varepsilon/2)\mathsf{OPT} \leq -(1-\varepsilon)\mathsf{OPT} \enspace.$$

(c) Using the fact that $Ax^{(0)} - \mathbb{1} \leq -\varepsilon/2 \cdot \mathbb{1}$, we compute $f_\mu(x^{(0)})$ as follows:
$$f_\mu(x^{(0)}) = \mu \sum_j \exp^{\frac{1}{\mu}((Ax^{(0)})_j - 1)} - \mathbb{1}^T x^{(0)} \leq \mu \sum_j \exp^{\frac{-\varepsilon/2}{\mu}} - \frac{1-\varepsilon/2}{n} \leq \frac{\mu m}{(nm)^2} - \frac{1 - \varepsilon/2}{n} \leq -\frac{1-\varepsilon}{n} \enspace.$$
Above, we have used that $\mathbb{1}^T x^{(0)} \geq x_i^{(0)} = \frac{1-\varepsilon/2}{n}$, where $i$ is the column such that $\|A_{:i}\|_\infty = 1$.

(d) To show $Ax \leq (1+\varepsilon)\mathbb{1}$, we can assume that $v = \max_j((Ax)_j - 1) \geq 0$ because otherwise we are done. Under this definition, we have $Ax \leq (1+v)\mathbb{1}$ and therefore $\mathbb{1}^T x \leq (1+v)\mathsf{OPT}$ by the definition of $\mathsf{OPT}$. We compute $f_\mu(x)$ as follows.
$$f_\mu(x) \geq \mu \exp^{\frac{v}{\mu}} - (1+v)\mathsf{OPT} \geq \mu\left(\left(\frac{nm}{\varepsilon}\right)^4\right)^{v/\varepsilon} - (1+v)n = \frac{\varepsilon}{4 \log(nm/\varepsilon)}\left(\left(\frac{nm}{\varepsilon}\right)^4\right)^{v/\varepsilon} - (1+v)n \enspace.$$
It is easy to see that the above quantity is positive whenever $v \geq \varepsilon$, and therefore, to satisfy $f_\mu(x) \leq 0$ we must have $v \leq \varepsilon$, which is equivalent to $Ax \leq (1+\varepsilon)\mathbb{1}$.

Finally, we notice that $Ax \leq (1+\varepsilon)\mathbb{1}$ implies $\mathbb{1}^T x \leq (1+\varepsilon)\mathsf{OPT}$ by the definition of $\mathsf{OPT}$.

(e) For any $x$ satisfying $f_\mu(x) \leq -(1 - O(\varepsilon))\mathsf{OPT} \leq 0$, owing to Proposition 2.2.d, we first have that $x$ is approximately feasible, i.e., $Ax \leq (1+\varepsilon)\mathbb{1}$. Next, because $-\mathbb{1}^T x \leq f_\mu(x) \leq -(1-O(\varepsilon))\mathsf{OPT}$, we know that $x$ yields an objective $\mathbb{1}^T x \geq (1-O(\varepsilon))\mathsf{OPT}$. Letting $x' = \frac{1}{1+\varepsilon}x$, we both have that $x'$ is feasible (i.e., $Ax' \leq \mathbb{1}$), and $x'$ has an objective $\mathbb{1}^T x'$ at least as large as $(1 - O(\varepsilon))\mathsf{OPT}$.

(f) Straightforward by some simple computation. □

## D Parallelizable Covering LP Solver

We divide our results on the covering LP into two parts. In the first part (see Section D.1), we show that the objective $\mathbb{1}^T \overline{y}$ is close to $\mathsf{OPT}$; in the second part (see Section D.2), we show that $A^T \overline{y} \geq (1 - 2\varepsilon)\mathbb{1}$ is approximately feasible. Both of our two steps rely on (4.1).

## D.1 Objective Optimality

We now show that the covering LP objective $\mathbb{1}^T \overline{y} \leq (1 + O(\varepsilon))\mathsf{OPT}$ as long as $T \geq \Omega(\frac{\log(nm/\varepsilon)}{\varepsilon^3})$. Note that this is smaller than that of $T \geq \Omega(\frac{\log n \cdot \log(nm/\varepsilon)}{\varepsilon^3})$ required in Theorem 3.5; however, as we shall see, it does not imply a faster convergence rate for covering LP than packing LP, because obtaining the approximate feasibility (i.e., $A^T \overline{y} \geq (1 - 2\varepsilon)\mathsf{OPT}$) requires more iterations.

The following lemma can be deduced essentially by (1) substituting $u = 0$ into (4.1), and (2) noticing that $\langle \nabla f_\mu(x^{(k)}), x^{(k)} \rangle \approx \mathbb{1}^T y(x^{(k)}) - \mathbb{1}^T x^{(k)}$ is approximately the duality gap at step $k$.

**Lemma D.1.** *For any $T \geq \frac{6}{\alpha\varepsilon} = \Omega(\frac{\log(nm/\varepsilon)}{\varepsilon^3})$, we have that $\mathbb{1}^T \overline{y} \leq (1 + 5\varepsilon)\mathsf{OPT}$.*

*Proof.* Substituting $u = 0$ into inequality (4.1), and using the fact that $V_{x^{(0)}}(0) = \mathbb{1}^T x^{(0)} \leq 1$, we obtain

$$\frac{1}{T} \sum_{k=0}^{T-1} \langle \nabla f_\mu(x^{(k)}), x^{(k)} \rangle \leq \frac{4}{\alpha T}(f_\mu(x^{(0)}) - f_\mu(x^{(T)})) + \frac{1}{\alpha T} + 2\varepsilon\mathsf{OPT} \tag{D.1}$$

We now respectively lower and upper bound the two sides of (D.1) as follows. One one hand, using the definition of gradient, the left hand side of (D.1) is lower bounded as

$$\langle \nabla f_\mu(x^{(k)}), x^{(k)} \rangle = \langle A^T y(x^{(k)}), x^{(k)} \rangle - \mathbb{1}^T x^{(k)} = \langle y(x^{(k)}), Ax^{(k)} \rangle - \mathbb{1}^T x^{(k)}$$

$$= \sum_j \exp^{\frac{1}{\mu}((Ax^{(k)})_j - 1)} \cdot (Ax^{(k)})_j - \mathbb{1}^T x^{(k)}$$

$$\geq (1-\varepsilon) \sum_j \exp^{\frac{1}{\mu}((Ax^{(k)})_j - 1)} - \mathbb{1}^T x^{(k)} - m \cdot (\frac{\varepsilon}{nm})^4$$

$$= (1-\varepsilon) \mathbb{1}^T y(x^{(k)}) - \mathbb{1}^T x^{(k)} - m \cdot (\frac{\varepsilon}{nm})^4 \ . \tag{D.2}$$

Here, the (only) inequality is because if $(Ax^{(k)})_j < 1 - \varepsilon$ for some constraint $j \in [m]$, the corresponding $\exp^{\frac{1}{\mu}((Ax^{(k)})_j - 1)} \leq \exp^{-\varepsilon/\mu} = (\frac{\varepsilon}{nm})^4$ is very small.

On the other hand, since $Ax^{(T)} \leq (1 + \varepsilon)\mathbb{1}$ by Proposition 2.2.d, we must have $\mathbb{1}^T x^{(T)} \leq (1 + \varepsilon)\mathsf{OPT}$, and thus $f_\mu(x^{(T)}) \geq 0 - (1 + \varepsilon)\mathsf{OPT}$. This gives an upper bound on the right hand side of (D.1) that is $\frac{4(1+\varepsilon)}{\alpha T}\mathsf{OPT} + \frac{1}{\alpha T} + 2\varepsilon\mathsf{OPT} \leq 3\varepsilon\mathsf{OPT}$, due to our choice of $T \geq \frac{6}{\alpha\varepsilon}$.

Together, we deduce from (D.1) that

$$(1 - \varepsilon) \frac{1}{T} \sum_k \left( \mathbb{1}^T y(x^{(k)}) - \mathbb{1}^T x^{(k)} \right) - m \cdot (\frac{\varepsilon}{nm})^4 \leq 3\varepsilon\mathsf{OPT}$$

$$\implies \mathbb{1}^T \left( \frac{1}{T} \sum_k y(x^{(k)}) \right) \leq \frac{1}{T} \sum_k \mathbb{1}^T x^{(k)} + 4\varepsilon\mathsf{OPT} \leq (1 + \varepsilon)\mathsf{OPT} + 4\varepsilon\mathsf{OPT} \ ,$$

where the last inequality is from $\mathbb{1}^T x^{(k)} \leq (1 + \varepsilon)\mathsf{OPT}$ for each $k$. □

## D.2 Approximate Feasibility

The approximate feasibility is tricker to prove. Indeed, the first proof to come to one's mind only implies that for $A^T \overline{y} \geq (1 - 2\varepsilon)\mathbb{1}$ for $T \geq \Omega(\frac{\log(n\rho) \log(nm/\varepsilon)}{\varepsilon^3})$. Here, $\rho$ is the largest entry of $A$ (i.e., the width). This bound on $T$ is slightly weaker than that in Theorem 3.5 because $\log(n\rho)$ may be larger than $\log(n)$. Fortunately, this loss can be avoided thanks to one of the two fixes below:

- WIDTH REDUCTION PRE-PROCESSING. One can modify the positive LPs to ensure $\rho =$

**Algorithm 2** FixCoord($A, \varepsilon, \overline{y}$)

---

**Input:** $A \in \mathbb{R}_{\geq 0}^{m \times n}$, $\varepsilon \in (0, 1/10]$, and $\overline{y} \in \mathbb{R}_{\geq 0}^m$.
**Output:** $y \in \mathbb{R}_{\geq 0}^m$ that satisfies $A^T y \geq \mathbb{1}$.
1: $\overline{y}' \leftarrow \overline{y}$.
2: **for all** $i$ such that $\lambda_i \stackrel{\text{def}}{=} (A^T \overline{y})_i - 1 + \varepsilon \leq -\varepsilon$ **do**
3: $\quad$ Let $j \in [m]$ be the largest entry in the $i$-th column, i.e., $A_{i,j} = \|A_{:i}\|_\infty$.
4: $\quad \overline{y}'_j \leftarrow \overline{y}'_j + \frac{-\lambda_i}{A_{i,j}}$.
5: **end for**
6: **return** $\frac{\overline{y}'}{1 - 2\varepsilon}$.

---

$n^{O(1)}$.[15] However, this modification requires some initialization which, if implemented, would make our algorithm not semi-stateless (see Appendix B).

- COORDINATE FIX POST-PROCESSING. We prove below that, for the same requirement on $T \geq \Omega(\frac{\log(n) \log(nm/\varepsilon)}{\varepsilon^3})$ as Theorem 3.5, although $A^T \overline{y}$ may be smaller than $1 - \varepsilon$ for some coordinate, one can safely raise some coordinates of $\overline{y}$ to obtain $A^T \overline{y}' \geq (1 - \varepsilon)\mathbb{1}$, without increasing $\mathbb{1}^T \overline{y}$ too much.

More specifically,

**Lemma D.2.** Let $\rho = \max_{i,j} |A_{i,j}|$, and $\overline{y} = \frac{1}{T} \sum_{k=0}^{T-1} y(x^{(k)})$.

- If $T \geq \max\{\frac{6}{\alpha \varepsilon}, \frac{\log(4n^2 \rho)}{\alpha \varepsilon}\} = \Omega\big(\frac{\log(n\rho) \log(nm/\varepsilon)}{\varepsilon^3}\big)$, we have $A^T \overline{y} \geq (1 - 2\varepsilon)\mathbb{1}$.

- If $T \geq \frac{6 \log(2n)}{\alpha \varepsilon} = \Omega(\frac{\log n \cdot \log(nm/\varepsilon)}{\varepsilon^3})$ (which is the same choice of $T$ in PosLPSolver($A, \varepsilon$)), there exists some simple fix $\overline{y}'$ from FixCoord($A, \varepsilon, \overline{y}$) (see Algorithm 2) satisfying
$$A^T \overline{y}' \geq (1 - 2\varepsilon)\mathbb{1} \quad \text{and} \quad \mathbb{1}^T \overline{y}' \leq \mathbb{1}^T \overline{y} + \varepsilon \mathsf{OPT} \ .$$

The proof of this lemma is involved, but has a clear high level intuition behind it.

We extract from (4.1) out only those terms that have $u$ in it, and rewrite (4.1) as follows: (here we have used the definition of $\nabla f_\mu(x^{(k)}) = A^T y(x^{(k)}) - \mathbb{1}$)

$$0 \leq \star + \frac{1}{\alpha T} V_{x^{(0)}}(u) + \langle A^T \overline{y} - \mathbb{1} + \varepsilon \mathbb{1}, u \rangle \ . \tag{D.3}$$

Now, suppose that $A^T \overline{y} \geq (1 - 2\varepsilon)\mathbb{1}$ is violated, there must exist some coordinate $i$ such that $(A^T \overline{y} - \mathbb{1} + \varepsilon \mathbb{1})_i < -\varepsilon$ is very negative. In such as case, we let $u_k = 0$ for every $k \neq i$, and use the choice $T \geq \Omega(\frac{\log(n\rho)}{\alpha \varepsilon})$. Inequality (D.3) is then simplified as $0 \leq \star + O(\frac{\varepsilon}{\log(n\rho)}) \cdot (u_i \log u_i - u_i) - \varepsilon \cdot u_i$. However, we can choose $u_i = (n\rho)^{\Omega(1)}$ to be very large, making the right hand side very negative. This contradicts to inequality (D.3), and thus finishes the proof of $A^T \overline{y} \geq (1 - 2\varepsilon)\mathbb{1}$ for the first half of the lemma.

To obtain the second half, it is first easy to see that FixCoord($A, \varepsilon, \overline{y}$) is computing some $\overline{y}'$ satisfying $A^T \overline{y}' \geq (1 - 2\varepsilon)\mathbb{1}$, because $\overline{y}'$ is so constructed to fix every violation of $A^T \overline{y} \geq (1 - 2\varepsilon)\mathbb{1}$. What is much harder to prove is that $\mathbb{1}^T \overline{y}' \approx \mathbb{1}^T \overline{y}$. In fact, this can be obtained, after some careful computation, from (D.3) again. This time, we carefully choose a different $u$: we identify *all* coordinates $i$ such that $(A^T \overline{y} - \mathbb{1} + \varepsilon \mathbb{1})_i < -\varepsilon$, and let $u_i$ be large on all of them.

---

[15]This can be done informally as follows. Within a single column of $A$, if the largest and smallest entries are off from either other by a factor more than $n^{\Omega(1)}$, the smallest entry can be replaced with zero without sacrificing too much accuracy. With this in mind, we can zero out "small" entries of each column. Next, we can similarly zero out "large" columns across all columns, and re-scale $A$ to get $\rho = n^{O(1)}$.

*Proof of Lemma D.2.* This time, we rewrite (4.1) as

$$\frac{1}{T}\sum_{k=0}^{T-1}\langle\nabla f_\mu(x^{(k)}), x^{(k)} - u\rangle \leq \frac{4}{\alpha T}(f_\mu(x^{(0)}) - f_\mu(x^{(T)})) + \frac{1}{\alpha T}V_{x^{(0)}}(u) + 2\varepsilon\mathsf{OPT} + \varepsilon\mathbb{1}^T u$$

$$\leq \frac{1}{\alpha T}V_{x^{(0)}}(u) + 3\varepsilon\mathsf{OPT} + \varepsilon\mathbb{1}^T u$$

where the last inequality comes from the fact that $\frac{4}{\alpha T}(f_\mu(x^{(0)}) - f_\mu(x^{(T)})) \leq \varepsilon\mathsf{OPT}$, which we have already used once in the proof of Lemma D.1. Let us define

$$\phi(u) \stackrel{\text{def}}{=} \frac{1}{\alpha T}V_{x^{(0)}}(u) + \frac{1}{T}\sum_{k=0}^{T-1}\langle\nabla f_\mu(x^{(k)}), u - x^{(k)}\rangle + \varepsilon\mathbb{1}^T u$$

and according to the inequality above we have $\phi(u) \geq -3\varepsilon\mathsf{OPT}$ for any $u \geq 0$.

**Proof of the First Half of the Lemma.** Recall from (D.2) that

$$\langle\nabla f_\mu(x^{(k)}), x^{(k)}\rangle \geq (1-\varepsilon)\mathbb{1}^T y(x^{(k)}) - \mathbb{1}^T x^{(k)} - m\cdot\left(\frac{\varepsilon}{nm}\right)^4 \geq -\mathbb{1}^T x^{(k)} - m\cdot\left(\frac{\varepsilon}{nm}\right)^4 \geq -(1+2\varepsilon)\mathsf{OPT}$$

and therefore

$$\frac{1}{\alpha T}V_{x^{(0)}}(u) + \langle A^T\overline{y} - \mathbb{1}, u\rangle = \phi(u) + \langle\nabla f_\mu(x^{(k)}), x^{(k)}\rangle \geq -3\varepsilon\mathsf{OPT} - (1+2\varepsilon)\mathsf{OPT} \geq -(1+5\varepsilon)n \ .$$

If there is some coordinate $i^*$ such that $v \stackrel{\text{def}}{=} (A^T\overline{y})_{i^*} - 1 + \varepsilon \leq -\varepsilon$, we substitute $u = (0,0,\ldots,x_{i^*}^{(0)}\cdot e^{-\alpha vT}, 0, \ldots, 0)$ where $u_{i^*} = x_{i^*}^{(0)}\cdot e^{-\alpha vT}$ into the above inequality, and we get

$$\frac{1}{\alpha T}\left(u_{i^*}\log\frac{u_{i^*}}{x_{i^*}^{(0)}} - u_{i^*} + \sum_i x_i^{(0)}\right) + v\cdot u_{i^*} \geq -(1+5\varepsilon)n \ .$$

Since the left hand side equals to $\frac{1}{\alpha T}\left(-u_{i^*} + \sum_i x_i^{(0)}\right)$ by our choice of $u_{i^*}$, we immediately obtain $-u_{i^*} \geq -(1+6\varepsilon)n\cdot\alpha T > -2n$ from it. Substituting in the definition of $u_{i^*} = x_{i^*}^{(0)}\cdot e^{-\alpha vT} \geq \frac{1/2}{n\|A_{:i}\|_\infty}\cdot e^{\alpha\varepsilon T}$, we conclude that $T < \frac{\log(4n^2\|A_{:i}\|_\infty)}{\alpha\varepsilon}$. However, this contradicts to our choice of $T \geq \frac{\log(4n^2\rho)}{\alpha\varepsilon}$. In other words, for $T \geq \max\{\frac{6}{\alpha\varepsilon}, \frac{\log(4n^2\rho)}{\alpha\varepsilon}\}$, we must have $(A^T\overline{y})_i - 1 + \varepsilon > -\varepsilon$ for all $i$, finishing the proof of $A^T\overline{y} \geq (1-2\varepsilon)\mathbb{1}$.

**Proof of the Second Half of the Lemma.** This time, using the definition of $\phi(u)$ and the convexity of $f_\mu(x)$, we obtain

$$-3\varepsilon\mathsf{OPT} \leq \phi(u) \leq \frac{1}{\alpha T}V_{x^{(0)}}(u) + \frac{1}{T}\sum_{k=0}^{T-1}(f_\mu(u) - f_\mu(x^{(k)})) \ .$$

From now on let us denote by $\widetilde{u} \stackrel{\text{def}}{=} (1-\varepsilon/2)x^*$. Recall that our earlier analysis yields the following:

- $f_\mu(\widetilde{u}) \leq -(1-\varepsilon)\mathsf{OPT}$ owing to Proposition 2.2.b;
- $f_\mu(x^{(k)}) \geq -(1+\varepsilon)\mathsf{OPT}$, owing to Proposition 2.2.d and $\mathbb{1}^T x^{(k)} \leq (1+\varepsilon)\mathsf{OPT}$; and
- $V_{x^{(0)}}(\widetilde{u}) \leq 2\mathsf{OPT}\cdot\log(2n)$, owing to (3.2).

Together, we obtain that

$$-3\varepsilon\mathsf{OPT} \leq \min_{u\geq 0}\phi(u) \leq \phi(\widetilde{u}) \leq \frac{1}{\alpha T}V_{x^{(0)}}(\widetilde{u}) + 2\varepsilon\mathsf{OPT} \leq 3\varepsilon\mathsf{OPT} \ . \tag{D.4}$$

where the last inequality is from our choice of $T \geq \frac{6\log(2n)}{\alpha\varepsilon}$.

Next we decompose $\phi(u)$ as follows. We let $\phi(u) = \sum_i \phi^i(u_i) + \phi^0$, where

$$\phi^i(u_i) \stackrel{\text{def}}{=} \frac{1}{\alpha T}\Big(u_i \log \frac{u_i}{x_i^{(0)}} + x_i^{(0)} - u_i\Big) + ((A^T \overline{y})_i - 1 + \varepsilon) \cdot u_i \qquad \text{and} \qquad \phi^0 \stackrel{\text{def}}{=} \frac{1}{T}\sum_{k=0}^{T-1}\langle \nabla f_\mu(x^{(k)}), -x^{(k)}\rangle$$

Let us denote by $\lambda \stackrel{\text{def}}{=} A^T\overline{y} - \mathbb{1} + \varepsilon\mathbb{1}$. Then, for each $i$ such that $\lambda_i \leq -\varepsilon$, we make the choice $u_i^* \stackrel{\text{def}}{=} x^{(0)} \cdot e^{-\alpha \lambda_i T}$; otherwise we choose $u_i^* = \widetilde{u}_i$.

Focusing on each $i$ such that $\lambda_i \leq -\varepsilon$, we have $\phi^i(u_i^*) = \frac{1}{\alpha T}(x_i^{(0)} - u_i^*)$ and $\phi^i(\widetilde{u}_i) \geq \lambda_i \widetilde{u}_i$. This gives a lower bound on their difference

$$\phi^i(\widetilde{u}_i) - \phi^i(u_i^*) \geq \frac{1}{\alpha T}(u_i^* - x_i^{(0)}) + \lambda_i \widetilde{u}_i \ .$$

Before continuing to prettify the right hand side, we make a technical observation. Letting $T_0 \stackrel{\text{def}}{=} \frac{6 \log(2n)}{\alpha \varepsilon}$ so that $T \geq T_0$, we have

$$u_i^* = x^{(0)} \cdot e^{-\alpha \lambda_i T} \geq \frac{1}{2n\|A_{:i}\|_\infty} \cdot \Big((e^{\alpha \varepsilon T_0})^{T/T_0}\Big)^{-\lambda_i/\varepsilon} \geq \frac{1}{\|A_{:i}\|_\infty}\Big(\big(\frac{1}{2n} \cdot e^{\alpha \varepsilon T_0}\big)^{T/T_0}\Big)^{-\lambda_i/\varepsilon}$$
$$\geq \frac{1}{\|A_{:i}\|_\infty}\Big((100n)^{T/T_0}\Big)^{-\lambda_i/\varepsilon} \ . \tag{D.5}$$

Therefore, the lower bound on $\phi^i(\widetilde{u}_i) - \phi^i(u_i^*)$ can be simplified as

$$\phi^i(\widetilde{u}_i) - \phi^i(u_i^*) \stackrel{①}{\geq} \frac{1}{\alpha T}u_i^* + \lambda_i \widetilde{u}_i - \frac{\varepsilon}{\|A_{:i}\|_\infty} \stackrel{②}{\geq} \frac{1}{\alpha T}\frac{1}{\|A_{:i}\|_\infty}\Big((100n)^{T/T_0}\Big)^{-\lambda_i/\varepsilon} + \lambda_i \widetilde{u}_i - \frac{\varepsilon}{\|A_{:i}\|_\infty}$$
$$\stackrel{③}{\geq} \frac{1}{\alpha T}\frac{1}{\|A_{:i}\|_\infty}\Big((100n)^{T/T_0}\Big)^{-\lambda_i/\varepsilon} + \frac{2\lambda_i}{\|A_{:i}\|_\infty}$$
$$\stackrel{④}{\geq} \frac{1}{\alpha T_0}\frac{1}{\|A_{:i}\|_\infty}(100n)^{-\lambda_i/\varepsilon} + \frac{2\lambda_i}{\|A_{:i}\|_\infty}$$
$$\stackrel{⑤}{\geq} \frac{1}{\alpha T_0}\frac{1}{\|A_{:i}\|_\infty}(100n)\frac{-\lambda_i}{\varepsilon} + \frac{2\lambda_i}{\|A_{:i}\|_\infty}$$
$$\stackrel{⑥}{\geq} \frac{-10\lambda_i}{\|A_{:i}\|_\infty} + \frac{2\lambda_i}{\|A_{:i}\|_\infty} \geq \frac{-8\lambda_i}{\|A_{:i}\|_\infty} \ .$$

Here ① is using the fact that $\frac{1}{\alpha T}x_i^{(0)} \leq \varepsilon \cdot \frac{1}{n\|A_{:i}\|_\infty}$. ② is using (D.5). ③ is using the fact that $\widetilde{u}_i \leq \frac{1}{\|A_{:i}\|_\infty}$ (due to the feasibility $Au \leq \mathbb{1}$) and $\lambda_i \leq -\varepsilon$. ④ is obtained by realizing that the left hand side of ④ is minimized, over all possible $T \geq T_0$, at $T = T_0$. ⑤ is obtained by realizing that $(100n)^t \geq (100n)t$ for any $t \geq 1$. ⑥ is by the definition of $T_0 = \frac{6\log(2n)}{\alpha\varepsilon}$.

Finally, we combine this with (D.4) and get

$$\sum_{i\,:\,\lambda_i \leq -\varepsilon} \frac{-8\lambda_i}{\|A_{:i}\|_\infty} \leq \sum_{i\,:\,\lambda_i \leq -\varepsilon} \phi^i(\widetilde{u}_i) - \phi^i(u_i^*) = \sum_{i \in [n]} \phi^i(\widetilde{u}_i) - \phi^i(u_i^*) \leq \phi(\widetilde{u}) - \min_{u \geq 0}\phi(u) \leq 6\varepsilon\mathsf{OPT}$$

and therefore

$$\sum_{i\,:\,\lambda_i \leq -\varepsilon} \frac{-\lambda_i}{\|A_{:i}\|_\infty} < \varepsilon\mathsf{OPT} \ . \tag{D.6}$$

Now we come to the last step of the lemma. For each coordinate $i$ such that $\lambda_i = (A^T\overline{y})_i - 1 + \varepsilon \leq -\varepsilon$, we find the corresponding $j$ where $A_{i,j} = \|A_{:i}\|_\infty$, and push $\overline{y}_j$ up by an additive amount of $\frac{-\lambda_i}{A_{i,j}}$. Letting $\overline{y}'$ be this new vector, we automatically have that $A^T\overline{y}' \geq (1 - 2\varepsilon)\mathbb{1}$, and moreover, $\mathbb{1}^T\overline{y}' - \mathbb{1}^T\overline{y} \leq \varepsilon\mathsf{OPT}$ due to (D.6). $\square$

It is now easy to see that Lemma D.1 and Lemma D.2 together imply that **Theorem D.3** (Covering LP). *For any $T \geq \max\{\frac{6}{\alpha\varepsilon}, \frac{\log(4n^2\rho)}{\alpha\varepsilon}\} = \Omega\big(\frac{\log(n\rho)\log(nm/\varepsilon)}{\varepsilon^3}\big)$, we have that $\frac{\overline{y}}{1-2\varepsilon}$ is a $(1+O(\varepsilon))$-approximate solution for the covering LP (1.2).*

*Alternatively, for any $T \geq \frac{6\log(2n)}{\alpha\varepsilon} = \Omega(\frac{\log n \cdot \log(nm/\varepsilon)}{\varepsilon^3})$, letting*
$$(x, \overline{y}) = \mathtt{PosLPSolver}(A, \varepsilon) \quad \text{and} \quad y = \mathtt{FixCoord}(A, \varepsilon, \overline{y}) \ ,$$
*we have that $y$ is a $(1+O(\varepsilon))$-approximate solution for the covering LP (1.2).*